\documentclass[useAMS,usenatbib]{mn2e}

\usepackage{multirow}
\usepackage{pifont}
\usepackage{lscape}
\usepackage{soul}
\usepackage{float}

\usepackage{upgreek,amssymb}
\usepackage{amsmath}
\usepackage{multicol}
\usepackage{subfig}

\usepackage{graphicx}
\usepackage[toc,page]{appendix}
\usepackage{enumerate}
\usepackage{xcolor}

\newcommand*{\bpass}{\textsc{bpass}v2\:}
\newcommand*{\altbpass}{\textbf{\sc{bpass\:}}}
\newcommand*{\reaper}{\textsc{reaper}\:}

\title[Neutron Star Kicks II]{Neutron Star Kicks II: Revision and further testing of the conservation of momentum ``kick'' model}
\author[J. C. Bray, J. J. Eldridge]{J. C. Bray$^{1}$ \thanks{E-mail: john.bray@auckland.ac.nz}, J. J. Eldridge$^{1}$ \\$^{1}$Department of Physics, University of Auckland, Private Bag 92019, Auckland, New Zealand}

\pagerange{\pageref{firstpage}--\pageref{lastpage}} \pubyear{2015}

\begin{document}
\maketitle
\label{firstpage}
\begin{abstract}
In \cite{RN391} we proposed a simple neutron star ``kick'' formula, $v_{\rm kick}=\alpha\,(M_{\rm ejecta}\,/\,M_{\rm remnant}) + \beta\,$ to explain the observed 2D velocities of young single neutron stars. Using this kick we found that there is no statistically significant preference for a kick orientation nor for any of the three initial mass function (IMF) slopes tested, and that populations including binary stars reproduced the kick distribution better than single star only populations. However, recent analysis by \cite{RN432}, prompted us to revisit our basic assumptions and our new analysis has led to revised ``best-fit'' kick values of $\alpha=100\, {\rm km\,s^{-1}}$ and $\beta=-170\,{\rm km\,s^{-1}}$.  The reduction of $\beta$ to a negative value is due to using the 2D observed kick velocity distribution rather than the modelled 3D velocity distribution for neutron stars (NS). To further test the validity of the new kick, we have created synthetic populations of runaway star and double neutron star (DNS) binaries at solar metallicity ($Z=0.020$) using our best-fit kick. We find our new kick values create runaway star velocities and DNS period distributions in agreement with the comparable observational distributions with only the DNS eccentricities in tension with the observations. From our DNS and BH-BH datasets we estimate a predicted DNS merger rate at solar metallicity of 3,864$^{+1,570}_{-2,371}$ Gpc$^{-3}$yr$^{-1}$ and a BH-BH merger rate of 5$^{+40}_{-1}$ Gpc$^{-3}$yr$^{-1}$. 
\end{abstract}

\section{Introduction}
The need to constrain the supernova kick has never been more important than it is today. In the age of gravitational wave astronomy, any meaningful reverse engineering from detection to progenitor requires a causal link between the progenitor stars and the kick \citep{RN489}. 

In \cite{RN391} (hereafter referred to as Paper I), we proposed a simple neutron star (NS) kick formula, $v_{\rm kick}=\alpha\,(M_{\rm ejecta}  / M_{\rm remnant}) + \beta\,$ to explain the velocity distribution of single NSs. We used a subset of the \cite{RN165} data to represent the velocity distribution for nearby single NSs less than 3 million years (Myrs) of age and found that the values that reproduce this distribution were $\alpha=70\, {\rm km\,s^{-1}}$ and $\beta=110\,{\rm km\,s^{-1}}$ for the observed two-dimensional (2D) velocities and $\alpha=70\, {\rm km\,s^{-1}}$ and $\beta=120\,{\rm km\,s^{-1}}$ for their inferred three-dimensional (3D) velocity distribution.

Subsequently, \cite{RN432} suggested there was no plausible explanation for such a high value of the velocity constant $\beta$, unless nonstandard physical processes were considered. This prompted us to revisit the assumptions used in the code and the mapping of the 3D to 2D velocities in particular. 

While the comparison of synthetic to observational data is never a straight-forward process, the comparison of 3D synthetic velocities to 2D velocities calculated from proper motion measurements presents a unique challenge. Adopting a deconvolution algorithm, \cite{RN165} inferred a 3D velocity distribution from a sample of 73 of their 2D and one-dimensional (1D) velocities. While this distribution is widely used by theorists, \cite{RN224} point out that according to the derived probability distribution, two of the 73 velocities used in the algorithm to select the distribution have a vanishing small probability of actually occurring. More recently \cite{RN478} carried out an analysis of proper motions of 28 pulsars using more accurate very long baseline array (VLBA) interferometry data and found a distribution using two Maxwellian distributions to be a better fit to their data, pointing out that the single Maxwellian derived by \cite{RN165} seriously underestimates the number of low-velocity pulsars. 


In light of these studies, we revisit and expand the kick analysis carried out in Paper I, evaluating larger progenitor and remnant datasets and re-examining our calculation of 2D velocities from their 3D counterparts. From this more comprehensive study we recalculate our best-fit $\alpha$ and $\beta$ combination. We find our new $\beta$ value has taken on a significant negative value and we examine how this negative value could arise, then to further validate our kick model, we consider other observational tests including the runaway star velocity distribution, the DNS properties and gravitational wave (GW) rates.


This paper is structured as follows; in Section 2 we detail the \textsc{reaper} (Remnant Ejecta And Progenitor Explosion Relationship) code modifications; in Section 3 we revisit the comparison of our single NS synthetic 2D velocities to the observations and our synthetic 3D velocities to the inferred \cite{RN165} distribution and again use the two-sample Kolmogorov-Smirnov (KS) test to identify the new best-fit $\alpha$ and $\beta$ values to reproduce the velocity distribution for single NSs less than 3 Myrs old; in Section 4, we investigate what physical conditions might give rise to our new best-fit kick values; in Section 5 we compare our best-fit synthetic 2D runaway velocity distribution, DNS eccentricity and period distributions to the observations as well as calculating the delay-time distributions and merger rates for DNS and BH-BH systems and compare these to the LIGO collaboration estimates; in Section 6 we summarise our results and present our conclusions.

\section{Code Modifications}
A comprehensive description of the \reaper code can be found in Paper I and a full description of the \bpass stellar evolution models can be found in \cite{RN490}. 

Below we outline the differences in the \reaper code used in this research compared to that used in Paper I. 

\subsection{Analysis of Stellar structure}
The original \textsc{reaper} code identified and analysed progenitors experiencing core-collapse supernovae in the \bpass stellar evolution models where the initial primary star mass was greater than 4~M$_\odot$. This lower limit was chosen to ensure the inclusion of core-collapse supernovae in systems where the two stars merged or where a star experienced mass-gain from its companion. Our original code did not consider ``low-mass'' or ``stripped envelope'' supernovae nor did it identify Type Ia supernovae events. These outcomes have been added in the revised code with all \textsc{bpass}v2 progenitor masses from 0.1~M$_\odot$ to 300~M$_\odot$ being analysed. We also now record all evolutionary outcomes in binary systems, including non-supernovae events such as white-dwarf\textendash white-dwarf (WD-WD) binary systems (see Figure~\ref{fig:A1} in Appendix \ref{A2} for a full list of the evolutionary outcomes from the code). 

As a result of including these lower mass stars we have changed from using a Salpeter IMF \citep{SALPETER} to using an IMF based on \cite{RN473} and \cite{RN477}. For consistency with previous work in the field \citep{RN286,RN472}, we use a power-law slope of -1.3 for initial masses between 0.1 and 0.5~M$_\odot$ and a slope of -2.35 for initial masses from 0.5 to 300~M$_\odot$. For this work the initial binary fraction we use is 100 per cent but we highlight that a number of the binary systems merge during the evolution of the primary star resulting in a population of $\sim$ 20 per cent of single stars $>$7~M$_\odot$ prior to core-collapse. Adding in the number wide binary systems which would most likely be observed as single star systems, our binary fraction prior to core collapse for massive stars is broadly in line with the multiplicity fraction found by Sana et al (2011) for O stars of $\sim$ 70 per cent. 

In Paper I, we identified stars experiencing a core-collapse supernovae as those who at the end of their \textsc{bpass}v2 evolution (i.e. at the end of carbon burning or the onset of neon ignition) had a total mass of greater than or equal to 2~M$_\odot$, and had a carbon and oxygen (CO) core mass greater than or equal to 1.4~M$_\odot$, (the CO core mass is taken as a proxy for the iron core mass). Note that the \textsc{bpass}v2 code evolves the primary star and assumes the secondary or companion star, remains on the main sequence throughout the evolution of the primary.

While researchers are broadly in agreement of the final conditions leading to core-collapse supernovae the same cannot be said for low-mass / stripped envelope or prompt Type Ia supernovae, that may also explode as in \cite{RN146}. We prefer the term ``low-mass'' supernovae to stripped envelope and ``ultra-striped'' supernovae and we will use the term low-mass supernovae in the remainder of this paper. In a similar vein, we identify runaway thermonuclear fusion resulting from the evolution of a star as prompt Type Ia supernovae to distinguish it from the alternative process of runaway thermonuclear fusion as a result of a WD accreting mass from a binary companion or as a result of two WDs merging.


While we acknowledge that our definitions for the conditions leading to these events may differ slightly from those of other researchers, the physics of these events is still relatively poorly understood and as a result each definition can be argued to be equally valid. The conditions we adopt for each supernova type are outlined below and shown in Table~1. These are applied to stars evolving both in binary systems and in isolation either as a result of a supernova in a merged binary system or a supernova in a runaway star.

The definition used in Paper I for a star to experience core-collapse is shown in the last line of Table 1, all other pathways are newly introduced in this work. 


\subsection{Assignment of Remnant and Supernovae types}
For stars with a final mass greater than or equal to 2~M$_\odot$ and with a CO, core mass greater than 1.3~M$_\odot$ but less than 1.4~M$_\odot$, we examine the helium (He) shell and oxygen and neon (ONe) core masses. We assume that half of the He shell mass will be accreted onto the CO core in subsequent burning so if the final mass of the CO core plus half of the He shell is greater than or equal to 1.4~M$_\odot$ we assume the star will undergo some from of supernova. To determine which type of supernova, we examine the ONe core mass of the model to determine if any carbon burning has occurred. If the ONe core mass is greater than or equal to 0.1~M$_\odot$ the star is assumed to develop a core which will subsequently collapse. If the mass of the CO core plus half of the He shell is greater than or equal to 1.4~M$_\odot$ but the ONe core mass is less than 0.1~M$_\odot$ then no significant carbon burning has occurred and we assume the stars core is too small to experience core-collapse but since it has grown beyond the Chandrasehkar limit, the star will explode. We assume this results in a prompt Type Ia supernova. If the CO core plus half of the He shell is less than 1.4~M$_\odot$ the star will not collapse and ends its evolution as a WD.

Where the final mass of the star is less than 2~M$_\odot$ and the CO core mass is greater than or equal to 1.4~M$_\odot$, we assume some form of core-collapse and again examine the ONe core mass to determine if the end result is a low-mass core-collapse or a prompt Type Ia supernova. If the ONe core mass is greater than or equal to 0.1~M$_\odot$ then significant carbon burning has occurred and we assume they develop a core which will subsequently collapse resulting in a low-mass core-collapse supernova, otherwise a prompt Type Ia supernova will occur. 

If the final mass of the star is greater than or equal to 2~M$_\odot$ but the CO core is less than or equal to 1.3~M$_\odot$ we assume the final fate of the star is a WD.

\subsection{Evolution of Primary Stars that end their \altbpass Evolution as White Dwarfs}
For primary stars identified at the end of \bpass evolution as WDs, we complete their evolution as follows; if the envelope has been lost, we set the final mass of the WD as the total mass of star at the end of its evolution, if the WD possesses a He envelope then we set the mass of the WD as the CO core mass and if the WD still has a H envelope then we assume the total final mass of WD is between the CO and CO + He core masses, assuming some dredge up will occur as the envelope is being lost.

The final mass assigned is checked to ensure no WDs have exceeded the Chandrasekhar mass and if so we remove these models from the dataset. We find only six of the 9,910 models fail using this criteria. We use these revised masses to calculate the final separation of the WD-companion star binaries. If the separation of the two stars is less than the sum of their radii, we assume a common envelope evolution and any mass loss from the primary {\it decreases} the separation. If the separation of the two stars is greater than the sum of their radii we assume a detached binary system and any mass loss from the primary {\it increases} the separation. We adopt a simple separation modification where the separation is either increased or reduced by the ratio of the initial to final masses multiplied by the initial separation. From these revised masses and separations we calculate the new period to allow identification of the corresponding WD-companion star system in the {\sc{bpass}}v2 secondary models which we then use to evolve the secondary star.


\subsection{Changes to Velocity Calculations}
In our analysis in Paper I, we adopted the opposite approach to \cite{RN165} and attempted to recreate the observed 2D velocities from our synthetic 3D velocities calculated as a result of the supernovae. The 3D velocities were obtained using the \textsc{bpass}v2 single-star and binary-star evolution models and the \reaper code utilising the Monte Carlo method of generating probability distributions from repeated random angle selections for the kick direction applied to compact remnants formed in supernovae. In our binary system models, for each $\alpha$ and $\beta$ pairing, we calculate the resultant single NS velocities for each of the 250 random kick angles selected. These provide the remnant and runaway or intact binary system velocities in terms of the cartesian components $x, y$ and $z$. For all objects, the 3D velocity is calculated using Pythagoras' theorem from these $x, y$ and $z$ velocities. 



\subsubsection{Calculation of 2D velocities from 3D spacial velocities}
There are two significant changes to the way the 2D velocities have been calculated compared to Paper I, firstly in our original code we calculated the final 2D and radial velocities from the 3D velocity vector using the sine and cosine of a randomly chosen angle between $0$ and $\pi/2$. We believe using this simple angle calculation resulted in a systematic underestimation of the transverse velocities which was then compensated for by a significant constant, or floor velocity, $\beta$. A better choice would have been to use the random isotropic angle selection criteria to view the 3D velocity vector, i.e. to calculate transverse and radial velocities using the sine and cosine of $\phi$ where $\phi=\arccos(2x-1)$ with $x$ being a random number between 0 and 1. However since we have the 3D velocity in cartesian components we now utilise two random {\it extrinsic} rotations between 0 and $\pi$ around the $x$ and $y$ axes to calculate our 2D and radial velocities. We have compared this method to the isotropic angle selection method outlined above and found no significant difference in the new best-fit $\alpha$ and $\beta$ values. 

Secondly, in Paper I  we treated the creation of each single NS as a unique event and assigned a different viewing angle for each. However, for binary systems, the orientation of the velocity of a single NS created by the secondary supernova is related to that created by the primary supernova. This is because at the time of the primary supernovae, we assume the binary orbits to have circularised and the angular momentum vectors of the two stars have aligned. For disrupted binaries, we expect the angular momentum vector of the runaway to be largely unaffected by the primary supernova and by it's subsequent isolated evolution. In cases where the binaries survive the primary supernovae, the angular momentum vector relationship of the secondary to the primary will also remain, (although in this case it will be modified by the orbital plane inclination gained as a result of the kick from the primary supernova).

In addition, the relatively low velocities of the runaways and the surviving binary systems as a result of the primary supernova mean the distance they will have travelled in comparison to their distance from Earth will be small. For these reasons, we believe that the 2D velocities of all objects formed from the same binary pair (i.e. from either the primary or secondary supernovae), are best represented by using the same viewing angle rather than assigning a completely new random angle for each event. 





\begin{table*}	
\caption{Final assignation of evolutionary end-points determined by the total final mass and relative $\rm M_{He}$\,core, $\rm M_{CO}$\,core and $\rm M_{One}$\,core of the evolved star.}
\begin{center}
\begin{tabular}{ c c c c c c}
\hline\hline
 Final mass  & $\rm M_{CO}$\,core& ($\rm M_{He}$\,core+$\rm M_{CO}$\,core)/2& $\rm M_{ONe}$\,core & Result\\ 
 \hline\hline
 $<2M_\odot$ & $<1.4M_\odot$ &  - & - & WD\\ 
  $\geqslant2M_\odot$ & $\leqslant1.3M_\odot$ & - & - &  WD\\ 
 $\geqslant2M_\odot$ & $1.3<$CO$<1.4M_\odot$ &  $<1.4M_\odot$ & - & WD\\ 
\hline
$<2M_\odot$ & $\geqslant1.4M_\odot$ &-& $<0.1M_\odot$ & Prompt Ia\\ 
$\geqslant2M_\odot$ & $1.3<$CO$<1.4M_\odot$& $\geqslant1.4M_\odot$ & $<0.1M_\odot$ & Prompt Ia\\
\hline
$<2M_\odot$ & $\geqslant1.4M_\odot$ &  & $\geqslant0.1M_\odot$ & Low-mass CC\\  
$\geqslant2M_\odot$ & $1.3<$CO$<1.4M_\odot$ &  $\geqslant1.4M_\odot$ & $\geqslant0.1M_\odot$ &  CC\\
\hline
$\geqslant2M_\odot$ & $\geqslant1.4M_\odot$ &- & - &CC\\  
\hline\hline
\end{tabular}
\\[1.5pt]
\textit{Note.} $M_\odot$=Solar mass; CO=Carbon and Oxygen; He=Helium; ONe=Oxygen and Neon; WD=white-dwarf; CC=core-collapse supernova; Ia=Type Ia supernova
\label{table:1}
\end{center}
\end{table*}

\subsubsection{Combination of Velocities from Primary and Secondary Supernovae} 
Where some form of core-collapse is identified, the remnant and runaway or intact binary system velocities are determined from calculations provided in \cite{RN146} and \cite{RN194} as in Paper I, providing the velocities in terms of the cartesian components $x, y$ and $z$. For primary stars experiencing a Type Ia supernova, the secondary is identified as a runaway star with its $x$ velocity set as the orbital velocity at the time of the primary supernova \citep{RN466}, and the $y$ and $z$ velocities are assumed to be equal to zero. For all objects, the 3D velocity is calculated using Pythagoras' theorem from these $x, y$ and $z$ velocities. 

Another change to the code is in the way velocities from the first and second supernovae are combined. Because we are running multiple angle simulations we have opted to simply combine the cartesian components of the velocities gained from the primary and secondary supernovae. We then calculate the 3D velocity by combining the resulting cartesian components. This is in contrast to Paper I where we multiplied the second velocity vector by $\pi/4$ and added this at 90 degrees to the primary 3D velocity vector, effectively averaging the secondary velocity.

For the special case of binaries surviving the primary supernova, to ensure the correct orientation of the secondary supernova kick and resulting velocity vectors, we calculate the inclination angle of the orbital plane resulting from the primary supernova using the prescription outlined in \cite{RN469} and modify the $\phi$ kick angle used in the \cite{RN194} calculations to incorporate this inclination. The cartesian velocity components of the secondary NS are calculated and then added to system velocity taking into account the binary orbital plane inclination angle. This is particularly important when analysing velocities from the pole and equator-centred kicks.

For runaways that subsequently experience a supernova, the velocity components resulting from the secondary  supernovae are calculated and then the primary and secondary cartesian velocity components added together to obtain the final cartesian coordinate velocities for the object. 

\begin{figure}
	\vspace{10mm}
	\hspace{-2mm}
		\includegraphics[scale=0.70]{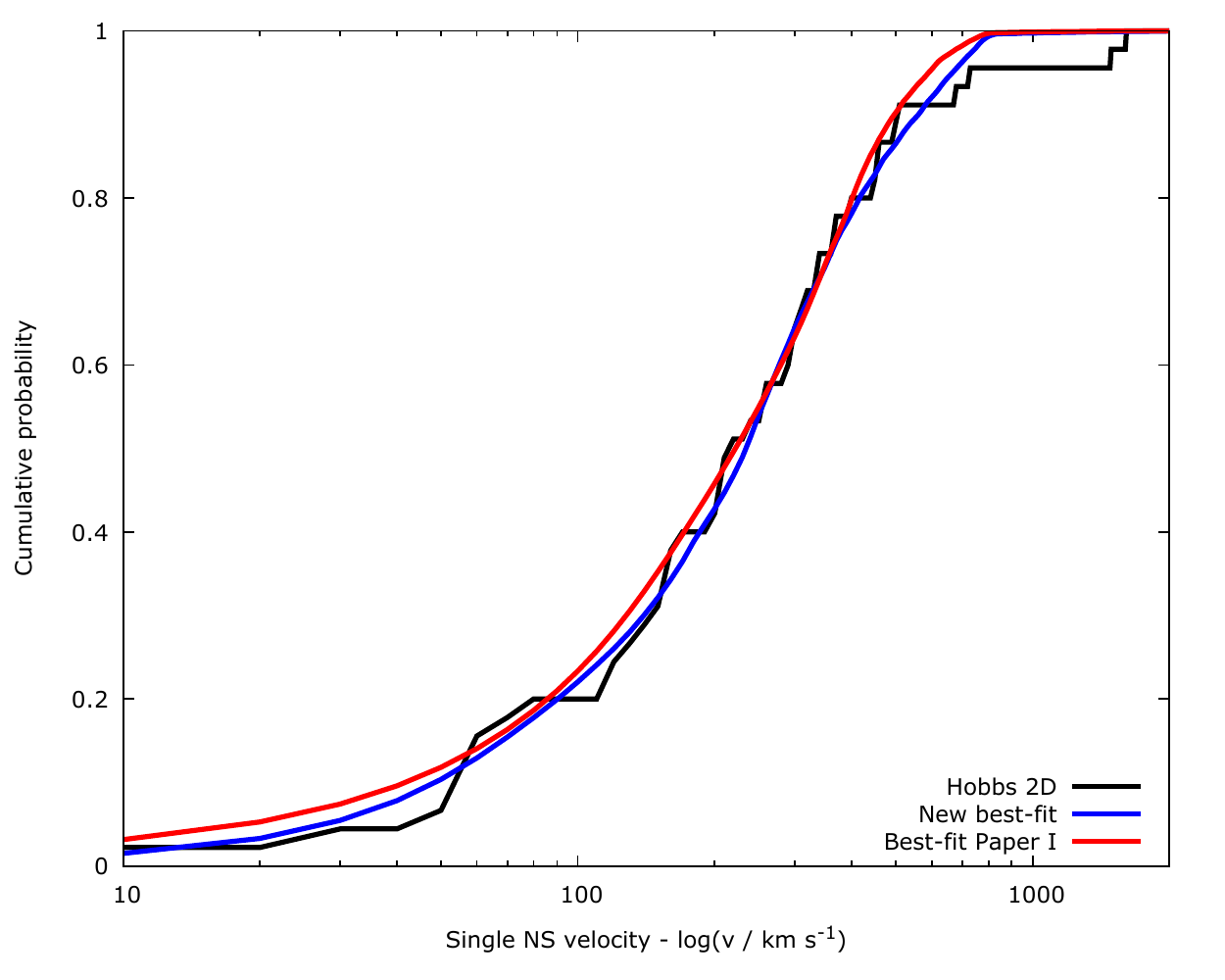}
  	\caption{Comparison of cumulative probability distributions for best fit of $\alpha$ and $\beta$ values to to the \protect\cite{RN165} two-dimensional (2D) velocity subset using an isotropic kick distribution - black; \protect\cite{RN165} subset, blue; New best-fit, red; Best-fit \protect\cite{RN391}.}
	\label{fig:newbestfit}
\end{figure}

The net effect of these revisions is subtle but visible in Figure~\ref{fig:newbestfit}, where probabilities for single NS velocities up to $\sim300\,{\rm km\,s^{-1}}$ have increased at the expense of higher velocity values. 

While these modifications have not changed the three fundamental results of Paper I, namely that there is no statistical preference for any of the three different kick orientations nor for any of the three different initial mass function (IMF) slopes and that populations including binary stars reproduced the kick distribution better than single star only populations, we believe the scale of the change to the best-fit $\beta$ value in particular requires further investigation which we detail in Section~4. 

\section{The new best-fit $\alpha$ and $\beta$ values : comparison of the synthetic to the observed single neutron star velocities}
As in Paper I we compare our single NS synthetic 3D velocities to the distribution derived by \cite{RN165} and our 2D distribution to our subset of the \cite{RN165} 2D velocities. In this work we use the same 2D velocity subset as that used in Paper I. This was obtained by selecting only those objects from the \cite{RN165} data that had proper motion measurements and characteristic ages less than 3 Myrs. We removed any pulsars known to be members of binary systems (see Section 2 of Paper I for rationale and more detail on single NS candidate selection). 

To ensure consistency with Paper I, we utilise the two-sample KS test to determine our best-fit $\alpha$ and $\beta$ values to these datasets and test an isotropic kick distribution as well as pole-centred and equator-centred distributions. Mimicking Paper I, we use the angles identified by \cite{RN203} for our maximum polar and equatorial kick angles. For the polar or spin-axis aligned kick, we utilise a maximum angle of $\pm$ 30$^{\circ}$ around the positive and negative $z$-axes, and for the equatorial or orthogonal to the spin-axis aligned kick, we use a maximum angle of $\pm$ 10$^{\circ}$ above and below the $xy$ plane. We use the arcos function outlined earlier to ensure an isotropic distribution within these ranges. For both the polar and equatorial kicks, because the spherical coordinate system used by \cite{RN146} and \cite{RN194} is centred around the $x$-axis, we create distributions around the polar and equatorial regions using the standard spherical coordinate system centred around the $z$-axis and then carry out a transformation to convert these points to the corresponding $\theta$ and $\phi$ angles used by \cite{RN146} and \cite{RN194}. 


The code modifications were expected to have a minimal effect on the 3D velocities and this is reflected in the similarity of the new best-fit $\alpha$ and $\beta$ values to those obtained in Paper I ($\alpha$=70 and $\beta$=120);

\begin{equation}
v_{\rm kick3D}/{\, \rm km\,s^{-1}} = 60^{+20}_{-20}\,\left(\frac{M_{\rm ejecta}}{M_{\rm remnant}}\right) +130^{+80}_{-80}
\end{equation}

Rather than compare our 3D velocity distributions to the \cite{RN165} inferred distribution, from this point onwards we focus on the comparison of our 2D velocity distributions to the 2D observational data, where there was a significant change to the best-fit values and to the $\beta$ value in particular. These results are summarised in Table 2.

\begin{figure}
\vspace{-10mm}
	\hspace{-6mm}
	\vspace{-5mm}
		\includegraphics[scale=0.80]{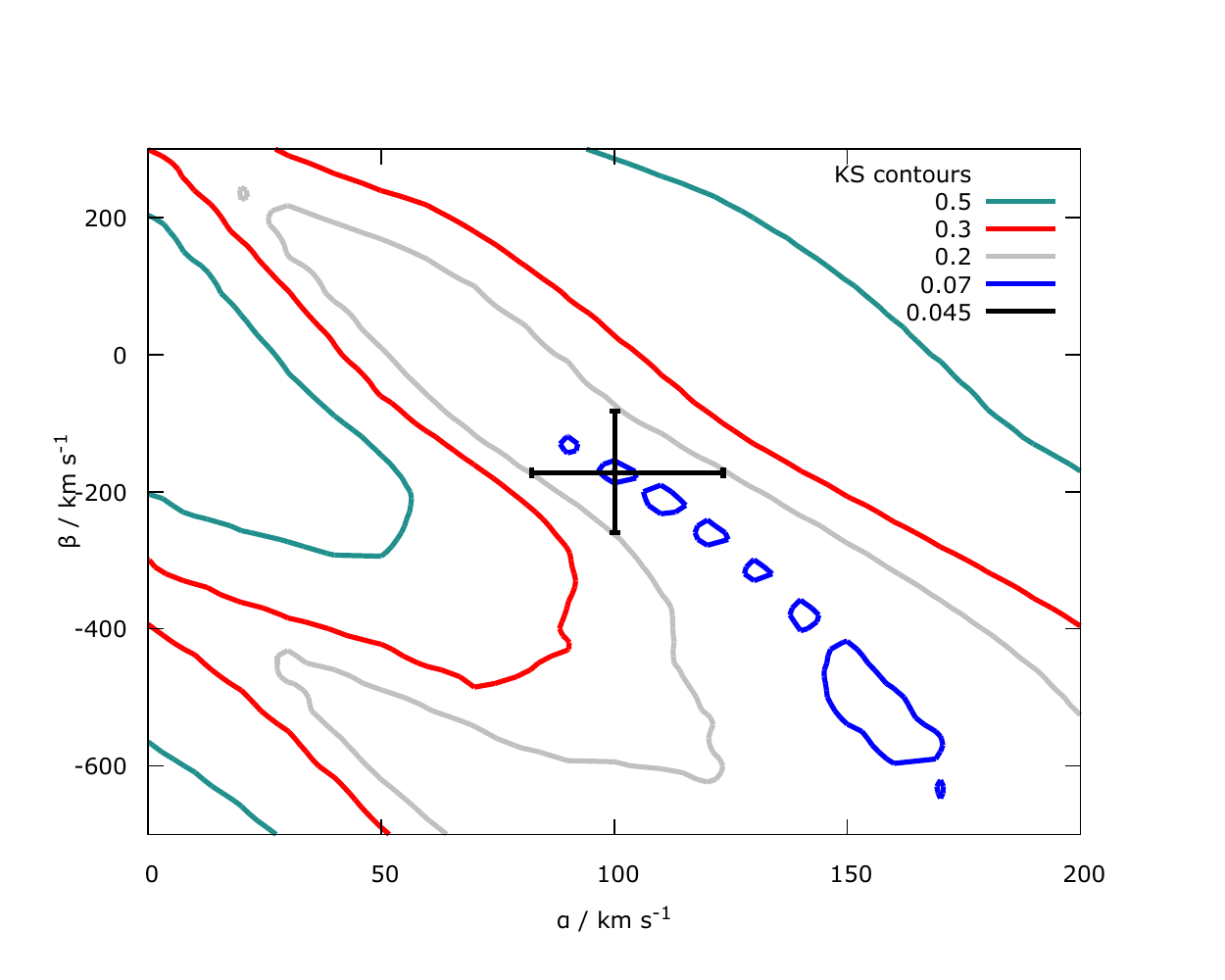}
  	\caption{Contour plot of two-sample Kolmogorov-Smirnov (KS) statistic comparing synthetic $\alpha$ and $\beta$ values to the \protect\cite{RN165} two-dimensional (2D) single neutron-star (NS) velocity subset using an isotropic kick distribution. The black cross shows our best-fit values of $\alpha$=100 $\beta$= -170 with error bars.}
	\label{fig:newbestfitcon}
\end{figure}

For our \cite{RN165} 2D subset using an isotropic kick we find best-fit $\alpha$ and $\beta$ values of;
\begin{equation}
v_{\rm kick2D}/\,{\rm km\,s^{-1}}= 100^{+30}_{-20}\,\left(\frac{M_{\rm ejecta}}{M_{\rm remnant}}\right) -170^{+100}_{-100}
\end{equation}
The KS statistic contour plot for the isotropic kick for 2D single NS velocities by $\alpha$ and $\beta$ value is shown in Figure~\ref{fig:newbestfitcon}. 

For the pole-centred kicks, we find the best-fit $\alpha$ and $\beta$ values of;
\begin{equation}
v_{\rm kick2D}/\,{\rm km\,s^{-1}}= 100^{+30}_{-20}\,\left(\frac{M_{\rm ejecta}}{M_{\rm remnant}}\right) -190^{+80}_{-90}
\end{equation}


For the equator-centred kicks, we find the best-fit $\alpha$ and $\beta$ values of;
\begin{equation}
v_{\rm kick2D}/\, \rm km s^{-1} = 100^{+30}_{-20}\,\left(\frac{M_{\rm ejecta}}{M_{\rm remnant}}\right) -160^{+110}_{-90}
\end{equation}

As in Paper I, we obtain our uncertainties from the KS critical D$_{\alpha}$ values for the two-sample KS statistic (we note that there is no relationship between the KS D$_{\alpha}$ value and our $\alpha$ variable other than the unfortunate use of the same Greek letter). We calculate the uncertainty of our best-fit $\alpha$ and $\beta$ values as the points where the KS statistic for our $\alpha$ and $\beta$ combination drops to the next critical D$_{\alpha}$ value. We consider $\alpha$ values of 0.2, 0.15, 0.1, 0.05, 0.025, 0.01, 0.005 and 0.001, where the 0.2 value represents the most stringent test. Where the KS value for our best-fit $\alpha$ and $\beta$ combination exceeds the 0.2 value we set the uncertainty as the $\alpha$ and $\beta$ values where the 0.2 value is reached. In the case of the 2D single NS velocity the best-fit of $\alpha$=100 and $\beta$= -170 exceeded the 0.2 value and hence our uncertainties were obtained where our KS statistic dropped to the 0.2 value as shown in Figure~\ref{fig:newbestfitcon}.

Again for both our 2D and 3D distribution comparisons, we find there is no statistically significant preference for any of the three kick orientations selected, although the isotropic kick is now slightly favoured over the pole and equatorial kick distributions. Again we find no statistically significant preference for any of the three IMF slopes tested with $\Gamma$=-2.70 only slightly favoured over the $\Gamma$=-2.00 and $\Gamma$=-2.35 slopes. The models containing binary stars provided a better fit than the single star only models but again we note the preference is not statistically significant. 

\begin{table*}	
\caption{Best-fit $\alpha$ and $\beta$ values and uncertainties for each kick orientation using 2D single NS velocities.}
\begin{tabular}{ c c c c c c c c}
\hline\hline
kick orientation&Dataset& $\alpha$ & Lower limit & Upper limit & $\beta$ & Upper limit & Lower limit \\ 
&&(${\rm km\,s^{-1}}$)&(${\rm km\,s^{-1}}$)&(${\rm km\,s^{-1}}$)&(${\rm km\,s^{-1}}$)&(${\rm km\,s^{-1}}$)&(${\rm km\,s^{-1}}$)\\
\hline\hline
Isotropic kick&Single NS velocity 2D & 100 & 80 &130& -170&-70&-270\\
\hline
Pole-centred kick&Single NS velocity 2D&100&80&130&-190&-110&-280\\
\hline
Equator-centred kick&Single NS velocity 2D&100&80&130& -160&-50&-250\\
\hline\hline
\label{table:2}
\end{tabular}
\\[1.5pt]
\textit{Note.} NS=neutron star; 2D=two-dimensional; 3D=three-dimensional
\end{table*}

Of course our derived kick is sensitive to the pulsar velocity distribution used and it is likely that some additional pulsars have been discovered since the \cite{RN165} survey was carried out as well as potentially more accurate proper motion and distance measurements for our selected pulsar subset. In addition the characteristic age is calculated from the pulsar braking index following a simple power law which has been shown to not necessarily apply to all pulsars \citep[e.g.][]{vela}. To test the effect of the age calculation on the selection of our pulsar subset and the effect of possible improvements in both proper motion and distance measurements on the velocities of our pulsar subset, we have created five new pulsar velocity distributions for single pulsars with proper motion measurements and obtained the best-fit $\alpha$ and $\beta$ combinations for each new subset. Our new distributions consist of a subset of the original \cite{RN165} pulsars but with the age cutoff extended to \textless7 Myrs, our original subset from the \cite{RN165} pulsars (age \textless3 Myrs), but with velocities calculated using the radial distances obtained from the electron density measurements of \cite{Yao:2016aa}, the new subset of the original \cite{RN165} pulsars with the age cutoff extended to \textless7 Myrs but with velocities calculated using the radial distances obtained from the electron density measurements of \cite{Yao:2016aa}, an entirely new pulsar subset selected from the Australia telescope national facility (ATNF) pulsar catalogue \citep[][http://www.atnf.csiro.au/research/pulsar/psrcat]{PulsarCat} with characteristic ages \textless3 Myrs and finally a new pulsar subset selected from the ATNF pulsar catalogue with characteristic ages \textless7 Myrs. 

We find in all cases, except that of the original \cite{RN165} pulsars with the age cutoff extended to \textless7 Myrs but with velocities calculated using the radial distances obtained from the electron density measurements of \cite{Yao:2016aa}, the new best-fit $\alpha$ and $\beta$ variables are within the uncertainties calculated from our original pulsar subset showing that the 2D velocity results are not substantially altered by uncertainties in the the distances or characteristic ages of our pulsar subset (refer Appendix \ref{A2} for velocity distributions and best-fit $\alpha$ and $\beta$ combinations).

For the remainder of the comparisons in this paper we utilise the best-fit $\alpha$ and $\beta$ combination obtained from our original pulsar dataset. We do this for a number of reasons, firstly this enables a meaningful comparison between the results obtained from our first paper \citep{RN391} and this paper, secondly the kick distribution obtained by \cite{RN165} using their pulsar sample is widely used by other researchers in the field so this ensures our results are comparable, thirdly by using the same pulsar sample as other researchers we ensure that any selection effects in the \cite{RN165} sample do not influence our results and lastly the new best-fit $\alpha$ and $\beta$ variables have a higher negative $\beta$ value which would most likely result in a DNS merger rate higher than the current predictions. We believe a better understanding of the kick will only come from more NS velocity measurements and better constraints on the DNS merger rate.
 
\section{The feasibility of our best-fit $\alpha$ and $\beta$ values}

While it is not the purpose of this paper to explain the physical origin of the $\alpha$ and $\beta$ values, we believe the significant difference between our latest best-fit $\alpha$ and $\beta$ values, and those found in Paper I requires some investigation. 

In contrast to our initial positive $\beta$ value, the new negative $\beta$ value suggests, in addition to the conservation of momentum velocity obtained as a result of the asymmetric ejection of the envelope, there is some additional force {\it reducing} the NS velocity. 

In Paper I we did not investigate the possibility of negative $\beta$ values, but in light of our binary code modifications identifying a negative $\beta$ value as the best-fit for binary star progenitors we enlarged our parameter space for single star kicks and re-ran our single star code. We find single star progenitors also obtain a best-fit with a negative $\beta$ value. Modifying the single star code to match the revised binary code, (i.e. to include low-mass supernovae and an isotropic view angle), we obtain best-fit values of $\alpha=130$ and $\beta=-470$.

\begin{equation}
v_{\rm kick2D}/\, \rm km s^{-1} = 130^{+20}_{-10}\,\left(\frac{M_{\rm ejecta}}{M_{\rm remnant}}\right) -470^{+100}_{-130}
\end{equation}

Although the best-fit for single stars is improved with this negative $\beta$ value, the binary star systems still provide a better fit to the single NS velocity observations.

We believe the similarity of the best-fit values for single and binary stars systems supports our binary star code changes (specifically the new view angle calculation and the viewing of both the primary and secondary supernovae from the same angle), and provides a more accurate representation of the observed 2D velocities.

\cite{RN432} and others, \citep{RN370,RN235,RN243,RN222,RN237}, maintain that rather than conservation of momentum being the main source of the kick, this imparts only a low velocity to the NS. They calculate that the main contribution to the NS velocity is the gravitational attraction between the NS and the more slowly moving, smaller ejecta-mass which accelerates the NS away from the strongest region of the explosion. The so-called ``gravitational tugboat'' effect. \cite{RN235} found that in one of their 2D models, NSs with velocities of over 1,000 km s$^{-1}$ were created showing that velocities similar to those observed are able to be produced using this mechanism. This effect is clearly significant, and in the context of our most recent best-fit values, seems to provide a logical explanation for the negative $\beta$ value we obtain. 


We propose that a significant initial conservation of momentum velocity is imparted to the NS as a result of ejecta-mass and/or velocity asymmetries, and then the relative gravitational attraction of the NS to the disparate ejecta-masses, reduces the velocity resulting in the negative $\beta$ value we obtain. 
 
To investigate if the magnitude of our negative $\beta$ value could be explained by such an effect, we return to our kick formula:
\begin{equation}
\hspace{10mm}
v_{\rm kick}/\,\rm km\,s^{-1}= \alpha \,\left(\frac{M_{\rm ejecta}}{M_{\rm remnant}}\right) + \beta
\end{equation}
We know that the 3D NS velocity is approximated by $v_{\rm kick}$ \citep{devan} and that the 2D velocity is approximately equal to 3D velocity times $\rm{\pi/4}$ which gives:
\begin{equation}
\hspace{10mm}
v_{\rm 2D}/\,\rm km\,s^{-1}=\frac{\pi}{4}\left( \alpha \,\left(\frac{M_{\rm ejecta}}{M_{\rm remnant}}\right) + \beta\right)
\end{equation}
Rearranging gives:
 \begin{equation}
\hspace{10mm}
v_{\rm 2D}\left(\frac{4}{\pi}\right)/\,\rm km\,s^{-1}= \alpha \,\left(\frac{M_{\rm ejecta}}{M_{\rm remnant}}\right) + \beta
\end{equation}
 if we now consider the mean observed 2D NS velocity in our sample set of $307\,{\rm km\,s^{-1}}$, the mean NS mass of 1.4~M$_\odot$ and insert our best-fit $\alpha$ and $\beta$ values we obtain:
\begin{equation}
\hspace{10mm}
391= 100\,\left(\frac{M_{\rm ejecta}}{1.4M_\odot}\right) -170
\label{equa:tug}
\end{equation}
Rearranging gives a required total ejecta-mass, which we now call $\rm{M_{Ej}}$, of 7.854~M$_\odot$. 

This equates to a mass for our progenitor at core-collapse of $\sim$~9.3M$_\odot$. This agrees with the initial progenitor mass range calculated by \cite{RN140} of between 7.5~M$_\odot$ and 19~M$_\odot$ for the most common supernovae type, Type II.

Adding 170 to each side gives:
\begin{equation}
\hspace{10mm}
561= 100\,\left(\frac{M_{\rm Ej}}{1.4M_\odot}\right)
\label{equa:fintug}
\end{equation}

We note that this is the same form as the equation given in \cite{RN238};

\begin{equation}
v_{\rm ns}= (\alpha_{JM}\,v_{\rm ej}) \left(\frac{M_{\rm ej}}{M_{\rm ns}}\right)
\end{equation}

In Equation \ref{equa:tug} our $\alpha$ value of 100 km s$^{-1}$ represents their anisotropy factor ($\alpha_{JM}$) multiplied by the ejecta velocity. Initial ejecta velocities are thought to be in the range of 8,000 - 15,000 km s$^{-1}$ \citep{RN251,RN467}, with ejecta velocities for Casiopea A (Cas A) estimated to be at the upper end of this range at 15,000 km s$^{-1}$ \citep{RN269}. If we assume an ejecta velocity of 10,000 km s$^{-1}$ and an asymmetry factor of 1 percent then we obtain our $\alpha$ value of 100. 

Equations (9) and (10) require that, in the case of a NS with the mean observed 2D velocity of $307\,{\rm km\,s^{-1}}$, the initial 3D velocity of the NS, gained as a result of the conservation of momentum must equal $561\,{\rm km\,s^{-1}}$. The same mass and velocity asymmetries that give rise to this velocity must also slow the NS by gravitational attraction to a final 3D velocity of $391\,{\rm km\,s^{-1}}$. 

We have written a simple Python code utilising basic kinematic formulae to test this hypothesis. For simplicity we ignore any gravitational effects from the companion star in our calculations. This ``toy'' model assumes a total ejecta-mass of 7.854~M$_\odot$ and that at time t=0 s, the ejecta velocity is $10,000\,{\rm km\,s^{-1}}$, the initial velocity of the NS due to the conservation of momentum must equal $561\,{\rm km\,s^{-1}}$, and that the NS is positioned at the centre of the supernova, (i.e. it gains its velocity instantaneously). We initially divide the ejecta-mass into four equal segments and add any mass asymmetry onto one equatorial plane segment and any velocity asymmetry onto the opposite equatorial plane segment. For simplicity we assume a point source for the masses and ignore the effect of the polar mass segments. We calculate the gravitational effects of the large and small ejecta-masses on the NS and vice versa but ignore the gravitational effect of the ejecta-masses on each other. This should have little effect on the final result as the high relative velocity of the ejecta-masses (20,000 km s$^{-1}$) and their small relative mass differential ($\sim$1 percent), mean their gravitational influence will be very short-lived and their relative effect on the NS almost equal and opposite. 

Mathematically this is a more general form of Equation (11), where the asymmetries in both ejecta mass and ejecta velocities are separated, giving:
\begin{equation}
\rm{V_{3D}}= \left(\frac{\left(\frac{M_{\rm EjT}}{4}+\Delta M\right)V_{Ej} - \frac{M_{\rm EjT}}{4}\left(V_{Ej}+\Delta V_{Ej}\right)}{1.4}\right)
\label{eqn:gen}
\end{equation}

While exhaustive testing was not carried out, we were able to easily locate configurations of varying mass, velocity, and ejecta location asymmetries where both the original conservation of momentum 3D velocity ($561\,{\rm km\,s^{-1}}$) and the final required NS 3D velocity ($391\,{\rm km\,s^{-1}}$) were obtained. For ejecta asymmetries located equidistant from the NS, if we set an ejecta-mass asymmetry of 1 percent, no velocity asymmetry, and the centre of mass of each equatorial segment of the ejecta at 623.5 km, the NS velocity was reduced from $561\,{\rm km\,s^{-1}}$ to $391\,{\rm km\,s^{-1}}$ in approximately 5 seconds. This is broadly in line with the timeframe of ``several seconds'' outlined by \cite{RN235} for their acceleration of the NS. Other simulations of varying mass and velocity asymmetries gave comparable results with timeframes up to 30 seconds.


Our simple simulations show that while the NS has a significant initial velocity toward the smaller ejecta-mass, in the first second it is slowed dramatically by the gravitation attraction of the larger ejecta-mass. However, because the initial velocity of the NS is in the opposite direction to the larger ejecta-mass, the separation distance between the NS and the larger ejecta-mass increases much more rapidly than that of the NS to smaller ejecta-mass. We observe an acceleration inversion at the point where the relative separations mean the greatest gravitational force on the NS switches from that exerted by the larger ejecta-mass to that exerted by the smaller ejecta-mass. At this time the separations are still relatively small and the NS now begins to accelerate toward the smaller ejecta-mass. In two of our simulations we observed a second velocity inversion which occurs when the square of the distances becomes similar and the NS is now accelerated once more toward the larger ejecta-mass. As the distance increases still further the acceleration reduces and the velocity asymptotes to the final value. We speculate that our ``toy'' model differs from the gravitational tugboat scenario in that it assumes a relatively high mass asymmetry in the ejecta-mass ($\sim$1 percent), at the time of the re-energisation of the shock. In this simplistic scenario, the higher asymmetry means the gravitational attraction of the NS to the larger ejecta-mass is the dominant force at very early times and hence the NS velocity is initially reduced. 

We highlight that in our \reaper code we have not enforced that the overall kick magnitude must be positive. This means for low ejecta-masses it is possible for the kick to be very small or even negative due to the high negative value of the $\beta$ constant. The negative kick of these low-mass stars can be understood in terms of the gravitational attraction of the more massive ejecta. Returning to our simple Python program, we show that with an ejecta-mass of 1~M$_\odot$, no velocity asymmetry, and a mass asymmetry of 1 percent, the NS initially gains a modest (positive) velocity toward the lower ejecta-mass, as a result of conservation of momentum, but this is quickly reversed by the gravitational attraction of the larger ejecta-mass. Again we did not carry out exhaustive testing but assuming a closer initial ejecta-mass location of 300 km (which does not seem unreasonable given this star has a mass of only 25 percent of the star we used in our previous simulation) and the above asymmetry profile, the NS gains a total velocity of approximately -120 km s$^{-1}$, i.e. toward the larger ejecta-mass.

With the exception of the small number of progenitors experiencing negative kicks, we believe that conservation of momentum kick is the major contributor to the NS velocity with the gravitational effect moderating this velocity rather than enhancing it. We highlight that in contrast to \cite{RN432,RN235,RN222,RN237}, where the gravitational attraction of the slower moving ejecta {\it{accelerating}} the NS is the dominant effect (the so-called ``gravitational tugboat'' effect), our NS gains an initially higher velocity from the asymmetric mass and velocity of the ejecta by conservation of momentum, and it's velocity is subsequently \textit{reduced} by the larger gravitational attraction of the more massive ejecta. In effect a ``gravitational towboat''. We also note that other mechanisms have been proposed to explain the single NS velocities observed \citep[e.g.][]{RN483}

\section{The best-fit $\alpha$ and $\beta$ values : further comparisons of synthetic to observed distributions}
Having ascertained that our new best-fit $\alpha$ and $\beta$ values have a plausible physical explanation, we now compare other synthetic datasets created using these best-fit values to the corresponding observations. As in Paper I, we continue to use the two-sample KS test, this time to compare our synthetic runaway velocity, DNS period and DNS eccentricity distributions to the observations. In addition we use our DNS and BH-BH datasets to calculate delay time distributions and merger rates and compare these to the LIGO estimates and the rates obtained by \cite{RN464} for a Milky Way equivalent galaxy (MWEG).
 
\subsection{Runaway Star Velocity Distribution}
For our observed runaway velocity distribution we use the 192 runaways identified by \cite{RN131} by their 3D (v$_{\rm pec}$) velocities. Their 3D dataset contains a low-velocity cutoff of v$_{\rm pec}= 28\, \rm km\,s^{-1}$, but also includes stars with lower velocities if the velocity vectors of these stars are anomalous when compared with their O and B type (OB) star clusters or associations. 

We note a possible selection bias in the \cite{RN131} dataset against low-mass stars since the mass-luminosity relationship will mean these are more difficult to detect at larger distances. We tested this effect on our dataset by introducing a minimum mass cutoff for our runaways, however the effect was only significant when the cutoff was set above 4~M$_\odot$ and we felt such a high mass cutoff was unreasonable so no minimum runaway mass was enforced.

We carry out two comparisons, firstly we compare the entire \cite{RN131} v$_{\rm pec}$ dataset with our entire synthetic runaway velocity dataset, then we repeat the comparison introducing a velocity cutoff of v$_{\rm pec}>28\, \rm km\,s^{-1}$ for both our synthetic runaway dataset and that of \cite{RN131}. 

The definition of a runaway varies from study to study with some researchers using solely a velocity cutoff while other use velocity and location. This makes direct comparison of single star velocities from the binary supernovae scenario (BSS) alone problematic \citep{RN207}. We assume runaways from the dynamical ejection scenario (DES) generally produce low-velocity neutron stars as a result of their escaping from the high gravitational potential cluster regions where they are formed and assume that these will be included in the low-velocity group identified by \cite{RN491}. Hence a velocity cutoff of v$_{\rm pec}>$28 km s$^{-1}$ is adopted to exclude these.

To calculate our runaway star velocity probabilities we use the primary star progenitor probability and weight this by the time the companion star spends as a runaway. Where the companion star has had a mass gain of more than 5 percent the star is assumed to be rejuvenated and it's runaway lifetime is defined as it's full evolutionary lifetime starting at it's new increased mass. Where the secondary experiences mass-loss or a mass-gain of less than 5 percent, it's lifetime is defined as it's evolutionary lifetime at the original mass less the lifetime of the primary star.

\subsubsection{No minimum runaway velocity}
Our best-fit to the observed 2D runaway velocity distribution using the two-sample KS test was obtained with $\alpha=0$ and $\beta=0$. Our best-fit to the single NS velocity distribution of $\alpha=100$ and $\beta=-170$ does not lie within the region of the highest D$_{\alpha}$ value. The presence of a large number of low-velocity objects in their dataset constrains the kick to be of very low magnitude.



\begin{figure*}
	\vspace{3mm}
	\hspace{-90mm}
		\includegraphics[scale=0.625]{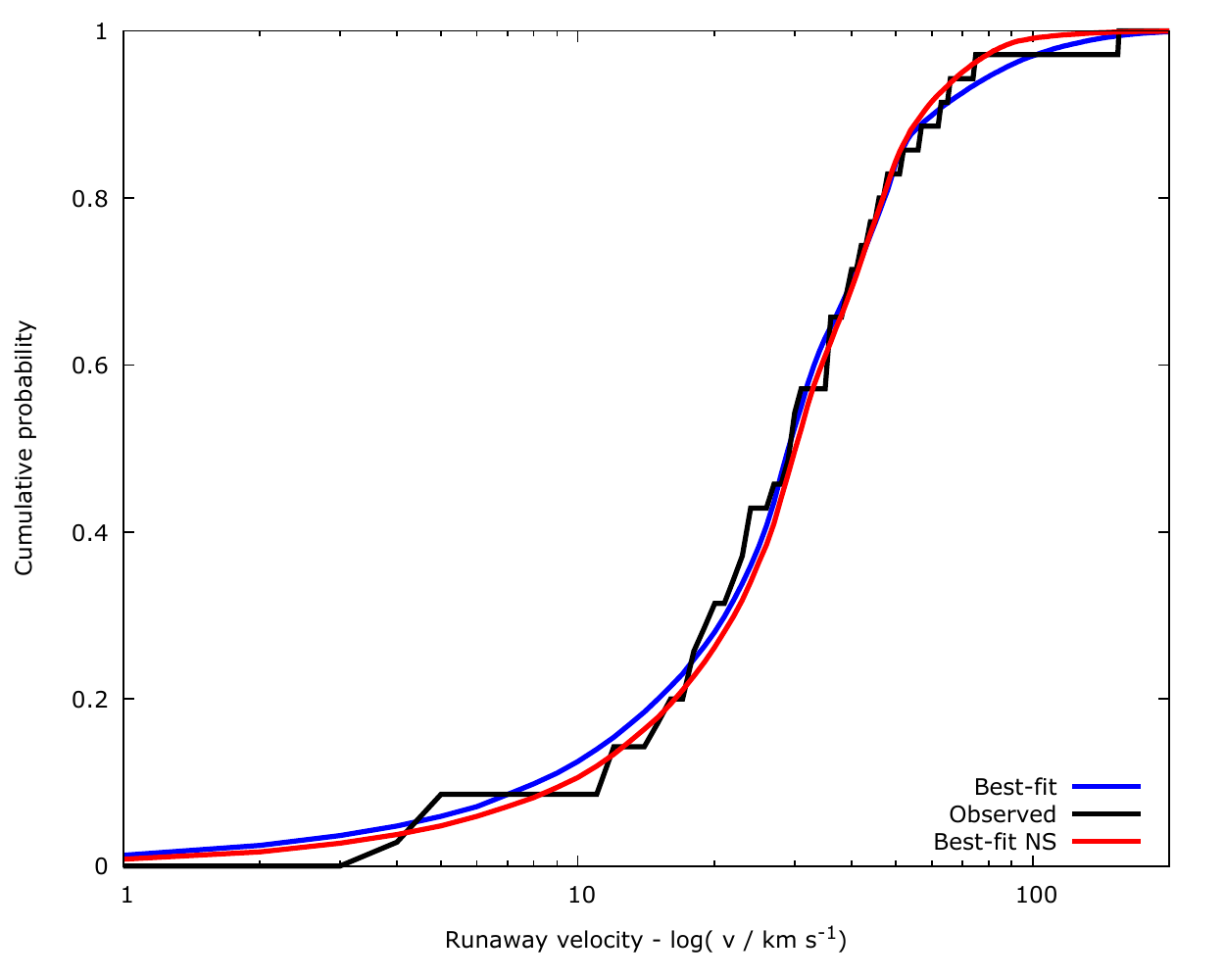}
\end{figure*}
\begin{figure*}
	\vspace{-77mm}
	\hspace{80mm}
		\includegraphics[scale=0.755]{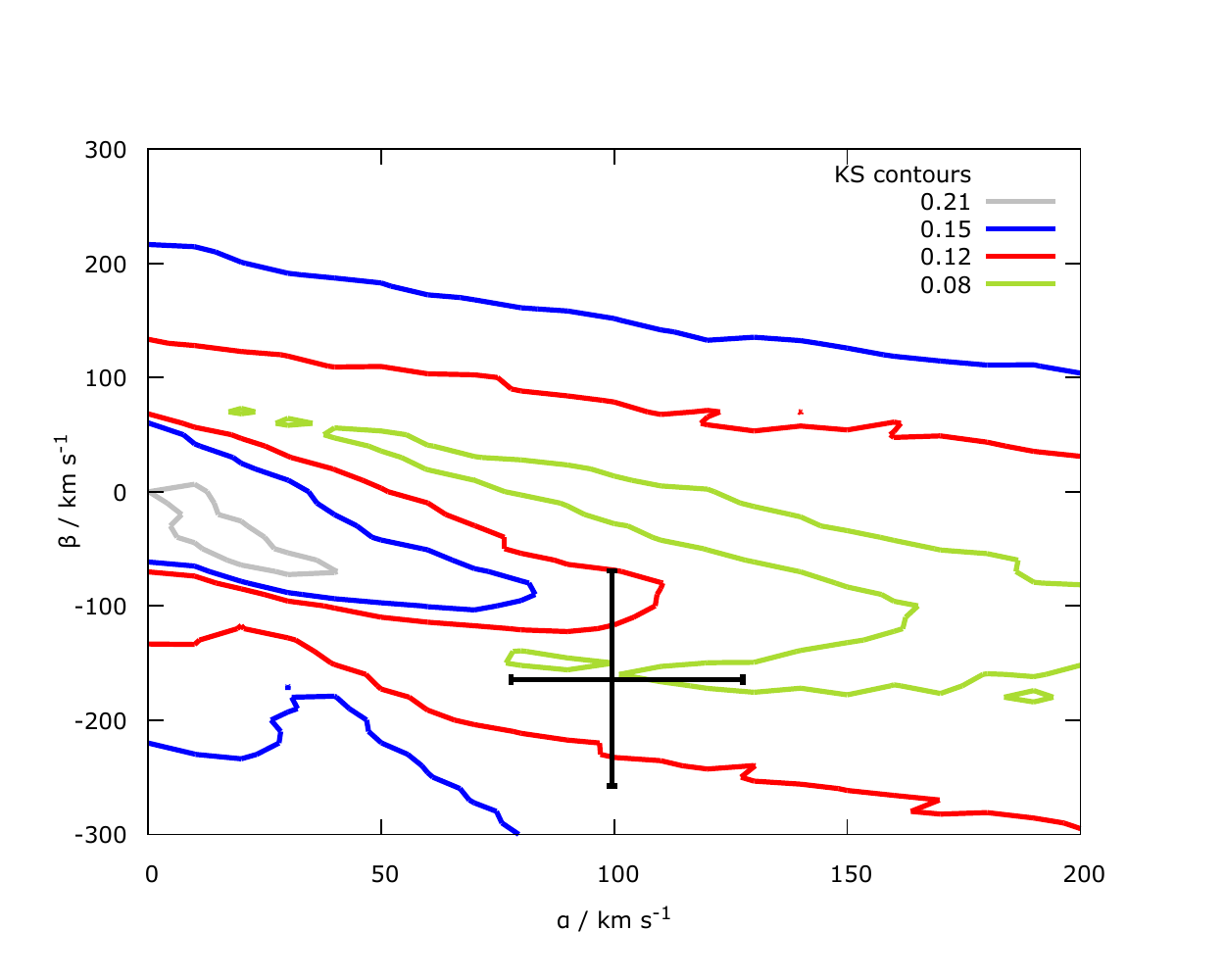}
	\caption{Comparison of synthetic runaway velocity distributions to the modified \protect\cite{RN131} 2D observational runaway velocity dataset, (v$_{\rm pec}>$28 km s$^{-1}$) using an isotropic kick distribution.\\ LH plot: Cumulative probability distribution - Blue - best-fit combination of $\alpha=110$ and $\beta=-20$ according to the two-sample Kolmogorov-Smirnov (KS) test: Black - 2D velocity observations (v$_{\rm pec}>$28 km s$^{-1}$); Red - best-fit to single neutron star (NS) velocity distribution of $\alpha=100$ and $\beta=-170$\\ RH plot: Contour plot of KS statistic for different $\alpha$ and $\beta$ pairings; The grey 0.21 contour line shows where the KS statistic drops to the next critical KS D$_{\alpha}$ value. The black cross shows our best-fit values of $\alpha=100$ and $\beta=-170$ with error bars.}
	\label{fig:rundat}
\end{figure*}

\subsubsection{Runaway velocity of v$_{\rm pec}>28\, \rm km\,s^{-1}$} 
Our best-fit to the observed 2D runaway velocity distribution, including only 3D velocities $>28\, \rm km\,s^{-1}$, using the two-sample KS test was obtained with $\alpha=110$ and $\beta=-20$ and is shown in the LH panel of Figure \ref{fig:rundat} along with the modified \cite{RN131} cumulative distribution and our best-fit $\alpha$ and $\beta$ values for our single NS velocities. The probability contour for our synthetic population by $\alpha$ and $\beta$ combination, is shown in the RH panel of Figure \ref{fig:rundat}. Our best-fit $\alpha$ and $\beta$ values for our single NS velocities with uncertainties added is shown as a black cross on the contour plot. 

After introducing the low-velocity cutoff the observed and best-fit synthetic runaway velocity distributions are almost indistinguishable.
The v$_{\rm pec}>28\, \rm km\,s^{-1}$ cutoff removes $\sim$82 percent of the observational dataset and $\sim$72 percent of the synthetic runaways. The difference suggests $\sim$15 percent of the \cite{RN131} are in fact from the DES. This is somewhat lower than the $\sim\frac{1}{3}$ suggested by \cite{RN262}.

\subsection{Comparison of Synthetic Data to Observations for DNS Binary Systems}
We use the 13 known DNS systems, as listed in Table \ref{table:dns} as our observational dataset. An additional system, B2127-11C was identified as being located in a globular cluster and was removed as the likelihood of the eccentricity and period being modified by interactions in such a dense environment was high.

For our eccentricity and period observations we include only the seven DNS systems in the bottom half of Table \ref{table:dns} that \cite{RN465} calculate do not merge in a time period of 50~Gyrs. We disregard the remainder as their periods and eccentricities may have been affected by gravitational wave emission. We make the same cutoff in our synthetic merger times only including those DNS systems that do not merge in a time period of 50~Gyrs.

Prior to our secondary supernovae analysis we removed any NS-companion star progenitor systems with orbital periods of less than or equal to five minutes ($\log$(P/days)$\leqslant$-2.5) as the \bpass secondary models do not include periods less than $\log$(P/days)=-2.0. These were assumed to result in the NS plunging into the companion star creating a Thorne-Zytkow object (TZO), before the occurrence of the secondary supernova. 

The remaining NS-companion star system periods were then rounded into bands of $\log$(P/days)=0.2, the NS remnant from the primary supernova set at 1.4~M$_\odot$ and the second star masses rounded to match the \bpass secondary model grid. Where a match is found, we take the evolved \bpass secondary models and analyse them in our \reaper code as before.

We make two different period comparisons. In the first, no upper limit was imposed on our synthetic periods, for our second comparison we introduce a cutoff of $\log$(P/days)$<$1.9 to match the maximum period of the observations. This period cutoff is implemented to remove any potential observational bias against observing long period DNS systems. 

\begin{table*}	
\caption{Properties of observed double neutron star (DNS) systems.}
\vspace{-2mm}
\begin{tabular}{c c c c c c}
\hline\hline
ID  & e & P & T$_{GW}$& Source used & Other sources \\ 
&&(days)&(Myrs)&object identified\\
\hline\hline
J0737-3039 & 0.088 & 0.102 &86&\cite{RN159}&\cite{RN139,RN364,RN438}\\
J1906+0746 & 0.085 & 0.166 &309&\cite{RN159}&\cite{RN364,RN438}\\
J1913+1102 & 0.09 & 0.206 &480&\cite{RN438}&\\
J1756-2251 & 0.181 & 0.32 &1660&\cite{RN159}&\cite{RN139,RN364,RN438}\\
B1913+16 & 0.617 & 0.323 &301&\cite{RN159}&\cite{RN139,RN364,RN438}\\
B1534+12 & 0.274 & 0.421 &2730&\cite{RN159}&\cite{RN139,RN364,RN438}\\
\hline
J1829+2456 & 0.139 & 1.176 &$\infty$& \cite{RN159} & \cite{RN139,RN364,RN438}\\
J0453+1559 & 0.113 & 4.072 &$\infty$& \cite{RN438} & \cite{RN139,RN364}\\
J1518+4904 & 0.249 & 8.634 &$\infty$& \cite{RN159} & \cite{RN139,RN364,RN438}\\
J1755-2550 & 0.089 & 9.696 &$\infty$& \cite{RN139} & \\
J1753-2240& 0.304 & 13.638 &$\infty$& \cite{RN159} & \cite{RN139,RN364,RN438}\\
J1811-1736 & 0.828 & 18.779 &$\infty$& \cite{RN159} & \cite{RN139,RN364,RN438}\\
J1930-1852 &  0.4 & 45.060 &$\infty$& \cite{RN438} & \cite{RN139,RN364}\\
\hline\hline
\label{table:dns}
\end{tabular}\vspace{-3mm}
\textit{Note.} ID=object identifier; e = eccentricity; P=orbital period; T$_{GW}$= evolution + merger time (T$_{GW}$ from \protect\cite{RN465})
\end{table*}

\begin{figure}
	\includegraphics[scale=0.65]{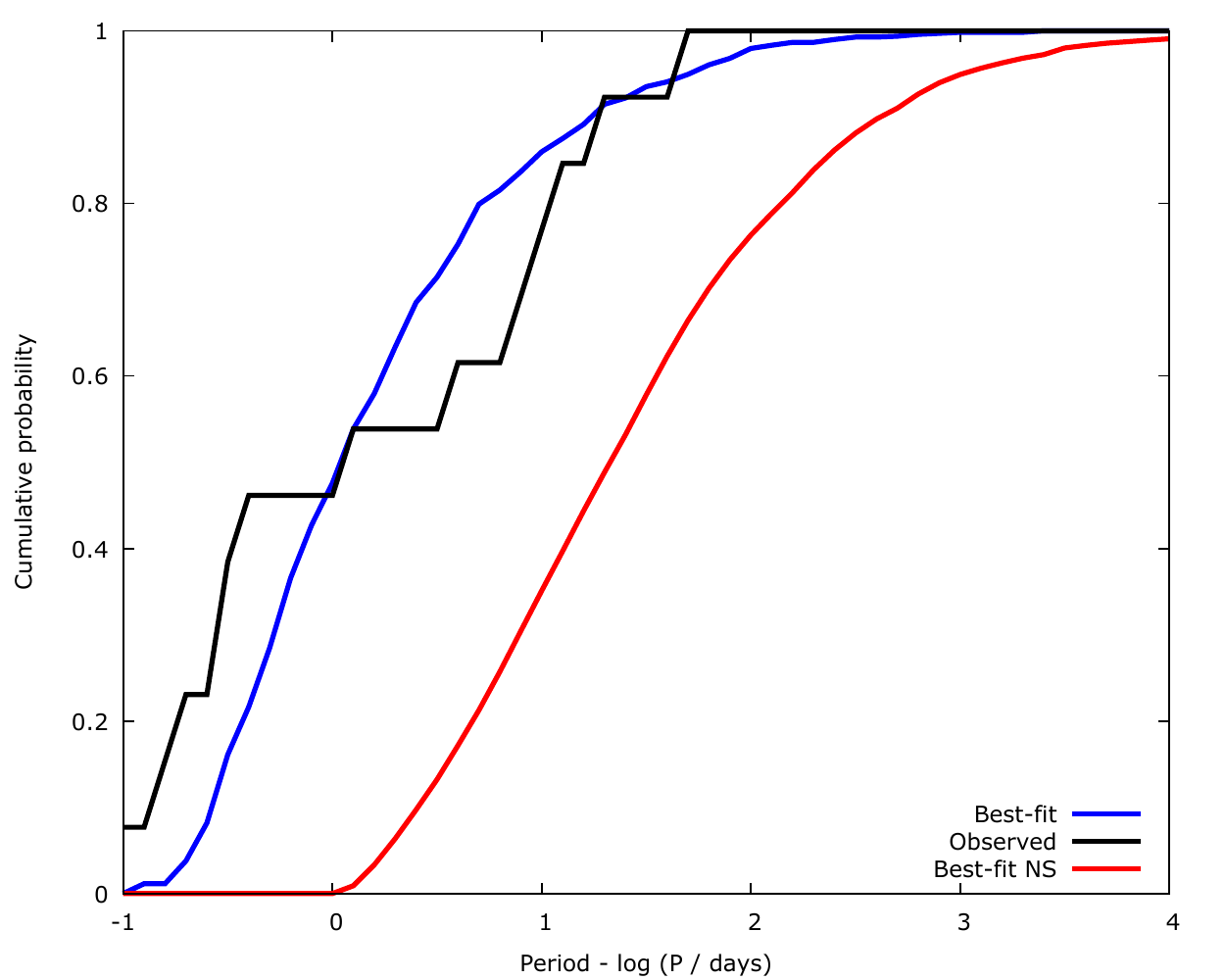}
	\caption{Comparison of synthetic double-neutron-star (DNS) periods to observed periods from Table~\ref{table:dns} using an isotropic kick distribution - no maximum synthetic period. Blue - best-fit combination of $\alpha=0$ and $\beta=-630$ according to the two-sample Kolmogorov-Smirnov (KS) test: Black - DNS period observations; Red - best-fit to single neutron star (NS) velocity dataset of $\alpha=100$ and $\beta=-170$}
	\label{fig:bfkper}
\end{figure}

\subsubsection{DNS period distribution}
{\bf{No period cutoff:}} The comparison of our synthetic to observed DNS period distribution, using no period cutoff, is achieved with $\alpha=0$ and $\beta=-630$ and is shown in Figure \ref{fig:bfkper} along with the observed cumulative period probability distribution and our best-fit kick of $\alpha=100$ and $\beta=-170$.

\begin{figure*}
	\vspace{-75mm}
	\hspace{81mm}
		\includegraphics[scale=0.75]{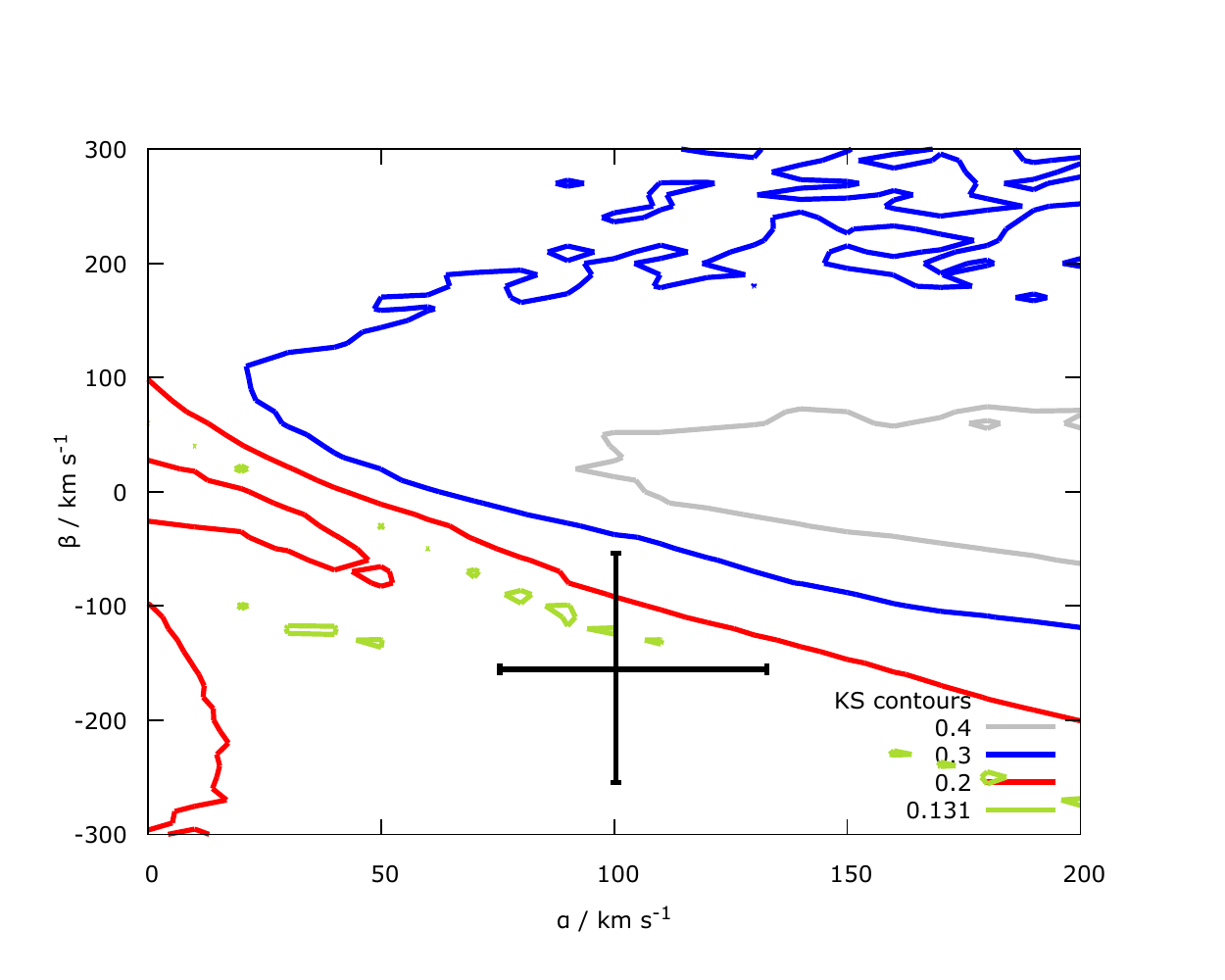}
	\caption{Comparison of synthetic double-neutron-star (DNS) periods to observed DNS periods from Table~\ref{table:dns} using an isotropic kick distribution - for systems with log (P/days)$<$1.9.\\ LH plot: Cumulative probability distributions. Blue - best-fit combination of $\alpha=20$ and $\beta=20$ according to the two-sample Kolmogorov-Smirnov (KS) test: Black - DNS period observations; Red - best-fit to single neutron star (NS) velocity distribution of $\alpha=100$ and $\beta=-170$\\ RH plot: Contour plot showing the two-sample KS statistic for $\alpha$ and $\beta$ combinations. The grey 0.4 contour line shows where the KS test statistic drops to the next critical KS D$_{\alpha}$ value. The black cross shows our best-fit values of $\alpha=100$ and $\beta=-170$ with error bars.}
	\label{fig:bfkperco}
\end{figure*}

{\bf{Period log (P/days)$<$1.9:}} The comparison of our synthetic to observed DNS period distribution, for systems with periods log (P/days)$<$1.9, is achieved with $\alpha=20$ and $\beta=20$. The cumulative probability distributions and the probability contour for our synthetic period distribution by $\alpha$ and $\beta$ combination is shown in Figure \ref{fig:bfkperco} along with the observed cumulative period probability distribution and our best-fit kick of $\alpha=100$ and $\beta=-170$.

The effect of removing systems with periods greater than 80 days is clearly evident, with the removal of the long period tail now meaning the synthetic period distribution is a very good fit to the observations. 
The period cutoff of log (P/days)$<$1.9 results in the removal of $\sim$40 percent of our synthetic DNS binaries thus we expect there may be at least six observable DNS systems in the Galaxy with periods of 80 days or longer.

\begin{figure*}
	\vspace{0mm}
	\hspace{-88mm}
		\includegraphics[scale=0.60]{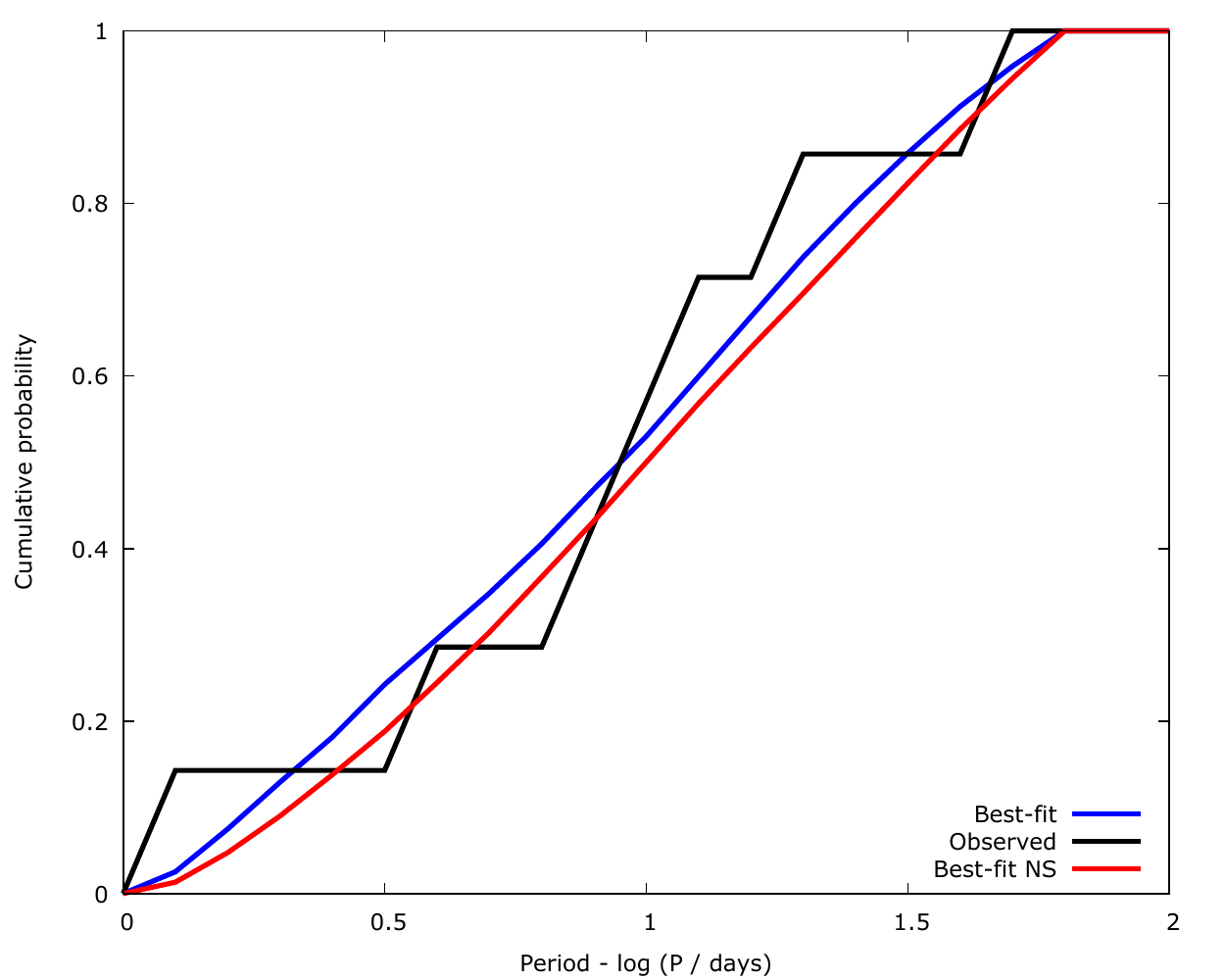}
\end{figure*}

\begin{figure*}
	\vspace{0mm}
	\hspace{-85mm}
		\includegraphics[scale=0.62]{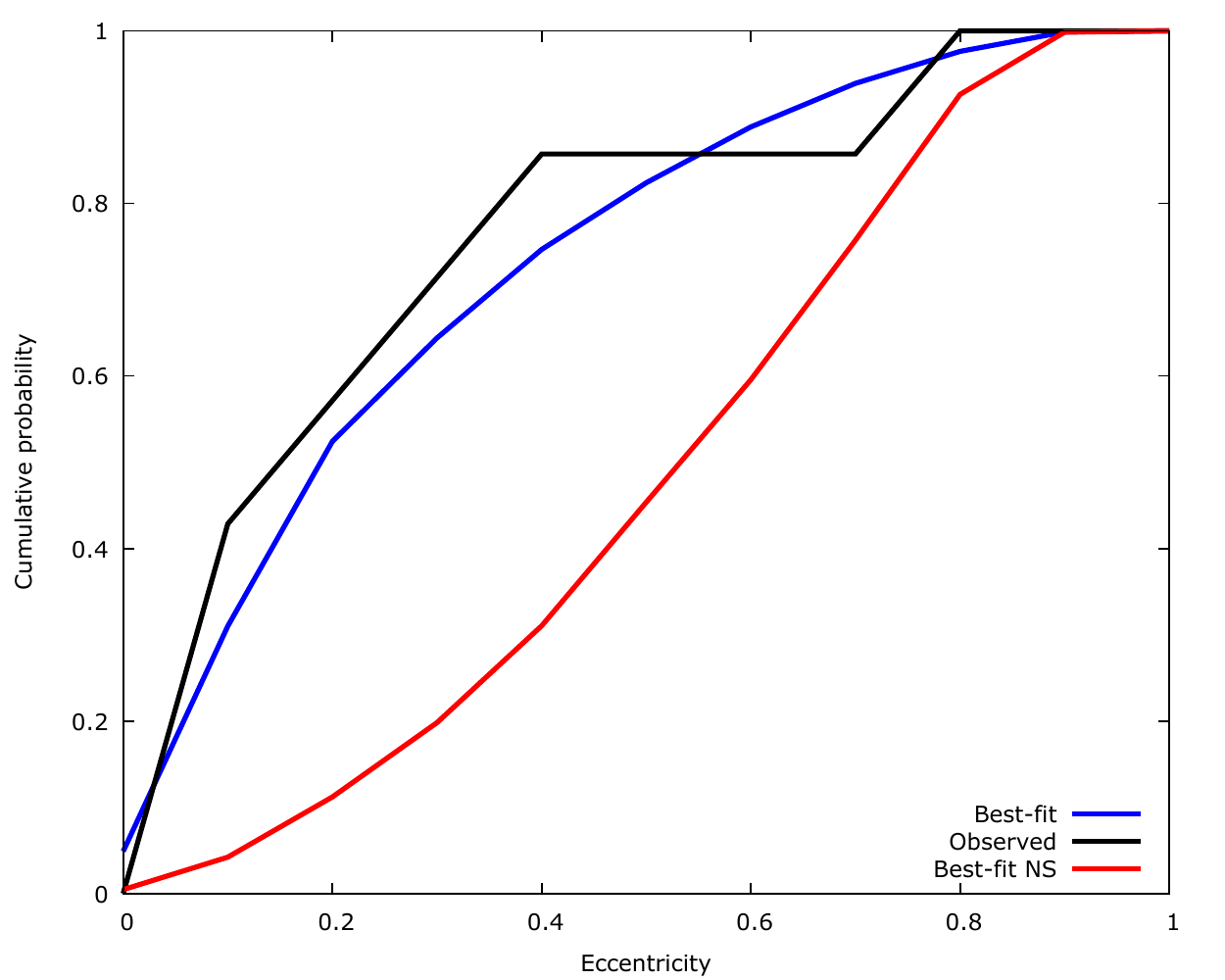}
\end{figure*}

\begin{figure*}
	\vspace{-77mm}
	\hspace{77mm}
		\includegraphics[scale=0.78]{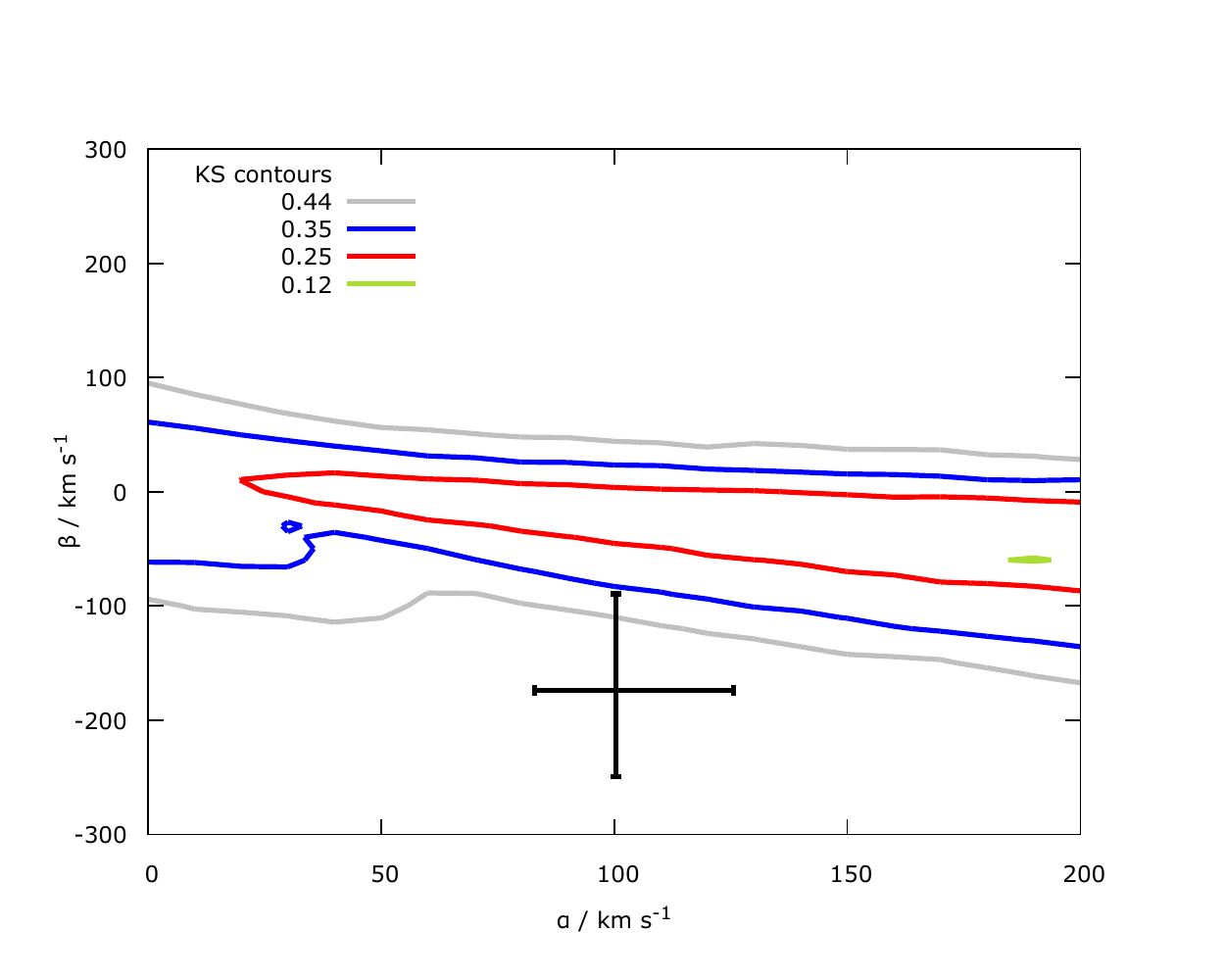}
	\caption{Comparison of synthetic DNS eccentricities to observed DNS eccentricities from Table \ref{table:dns} using an isotropic kick distribution - for systems with log (P/days)$<$1.9.\\ LH plot: Cumulative probability distribution. Blue - best-fit combination of $\alpha=190$ and $\beta=-60$ according to the two-sample Kolmogorov-Smirnov (KS) test: Black - DNS eccentricity observations; Red - best-fit to single neutron star (NS) velocity distribution of $\alpha=100$ and $\beta=-170$\\ RH plot: Contour plot showing KS statistic for $\alpha$ and $\beta$ combinations. The grey 0.44 contour line shows where the KS statistic drops to the next critical KS D$_{\alpha}$ value. The black cross shows our best-fit of $\alpha=100$ and $\beta=-170$ with error bars.}
	\label{fig:eccfit}
\end{figure*}

\subsubsection{DNS eccentricity distribution}

{\bf{No period cutoff:}} The best-fit of our synthetic to observed DNS eccentricity distribution was achieved with an $\alpha=190$ and $\beta=-60$.

{\bf{Period log (P/days)$<$1.9:}} The best-fit of our synthetic to observed DNS eccentricity distribution using log (P/days)$<$1.9 log days, is also achieved with $\alpha=190$ and $\beta=-60$ and is shown in the LH plot of Figure~\ref{fig:eccfit} along with the observed cumulative eccentricity probability distribution and our best-fit values for the single NS velocity distribution of $\alpha=100$ and $\beta=-170$. The contour plot by $\alpha$ and $\beta$ combination is shown in the RH plot of Figure \ref{fig:eccfit}.

We find our best-fit single NS values of $\alpha=100$ and $\beta=-170$ create systems significantly more eccentric than the observations, and while the uncertainty of our single NS best-fit values cross into the region of the highest D$_{\alpha}$ value, the observed eccentricities are not well reproduced by our synthetic best-fit values. 

\subsection{DNS Delay-time Distributions and Merger Rates}
To calculate our delay-time distributions we calculate the number of DNS mergers for an instantaneous starburst of $10^{6}$~M$_\odot$. We include the secondary star masses in our starburst mass and calculate the merger number per time-bin. We then divide the number in each time-bin by the time-bin width to obtain a merger rate. 

\begin{figure*}
	\hspace{-8mm}
		\includegraphics[scale=0.85]{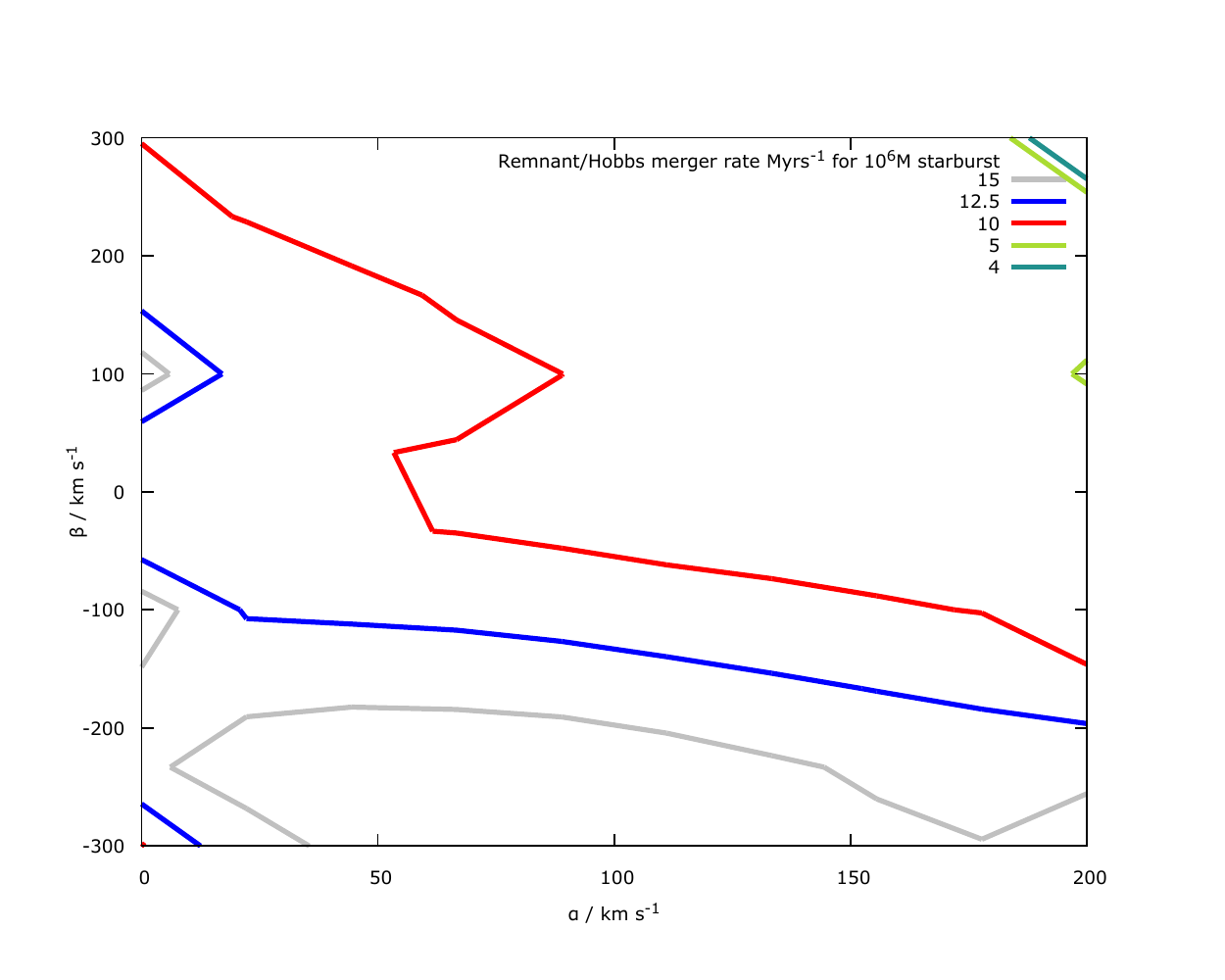}
  \caption{Contour plot of delay-time distribution for double-neutron-star (DNS) binary system mergers for our kick using different $\alpha$ and $\beta$ combinations divided by the delay-time distribution for double-neutron-star (DNS) binary system mergers using the \protect\cite{RN165} kick.} 
	\label{fig:cpremdivhobbs}
\end{figure*}

The effect on the merger rate for different $\alpha$ and $\beta$ combinations is shown in Figure \ref{fig:cpremdivhobbs}. The rate increase varies from nil, in the region of $\alpha>180$ and $\beta>250$ to 15 times the \cite{RN165} rate in the region where $\alpha\sim100$ and $\beta\sim-250$ km s$^{-1}$.

\begin{figure*}
	\vspace{-71mm}
	\hspace{77mm}
		\includegraphics[scale=0.65]{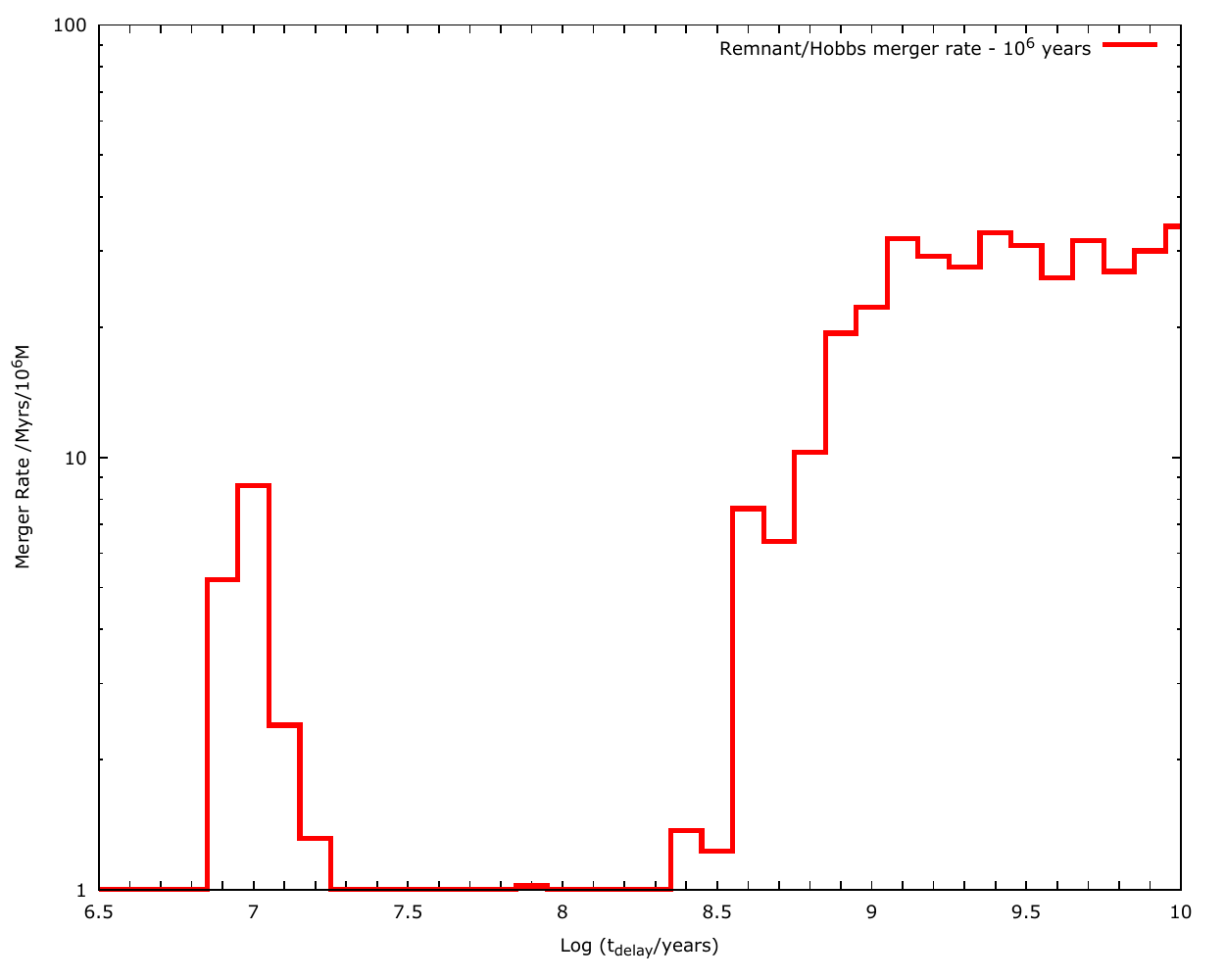}
	\caption{Delay-time distribution for double-neutron-star (DNS) mergers assuming an instantaneous starburst of $10^{6}$ M$_\odot$.\\ LH plot: Merger rate /Myrs/$10^6$M$_{\odot}$. Black; merger rate using \protect\cite{RN165} kick, Red; merger rate using $\alpha=100$ and $\beta=-170$.\\RH plot: Relative merger rate difference; best-fit merger rate / \protect\cite{RN165} merger rate.}
	\label{fig:dtd}
\end{figure*}

The comparison between the delay-time distribution created by the \cite{RN165} kick and our best-fit values of $\alpha=100$ and $\beta=-170$ is shown in the left-hand plot in Figure \ref{fig:dtd}. 


In simple terms our new best-fit kick variables of $\alpha=100$ and $\beta=-170$ generally result in our NSs receiving smaller magnitude kicks than they would using the \cite{RN165} kick distribution. As a result we obtain more intact DNS binaries and hence a higher number of mergers overall. 

To show the effect of the two different kick distributions on the merger rate Myrs$^{-1}$ for a starburst of $10^{6}$~M$_\odot$, in the RH plot of Figure \ref{fig:dtd} we show the merger rate using the best-fit kick ($\alpha=100$ and $\beta=-170$) divided by the merger rate using the \cite{RN165} kick distribution. 

We find the early time peak around $\log(t_{\rm delay}/\rm years)$=7, arises from the merger of non-rejuvenated secondaries with ejecta-masses $<$5.2~M$_\odot$. As a result of these relatively low ejecta-masses, these systems receive a secondary kick of $\sim$200\,km\,s$^{-1}$ which is $\sim$50\% of the average \cite{RN165} kick, hence the periods of these binaries are not increased as much by the secondary kick and they merge at earlier times. The increase in merger rates after $\log(t_{\rm delay}/\rm years)\sim$8.5 arises from rejuvenated secondaries with masses 10~M$_\odot<M<$15~M$_\odot$ in relatively short period systems ($0.5<\log(P/\rm days)<0.9$). These secondaries have very low ejecta-masses ($<$2~M$_\odot$) and as a result receive kicks $\sim$100\,km\,s$^{-1}$. We speculate that as a result of our very small kicks, these systems remain bound and merge at later times, whereas if they experienced a kick chosen from the \cite{RN165} kick distribution the majority would be disrupted. 

\begin{figure*}
	\vspace{-80mm}
	\hspace{81mm}
		\includegraphics[scale=0.75]{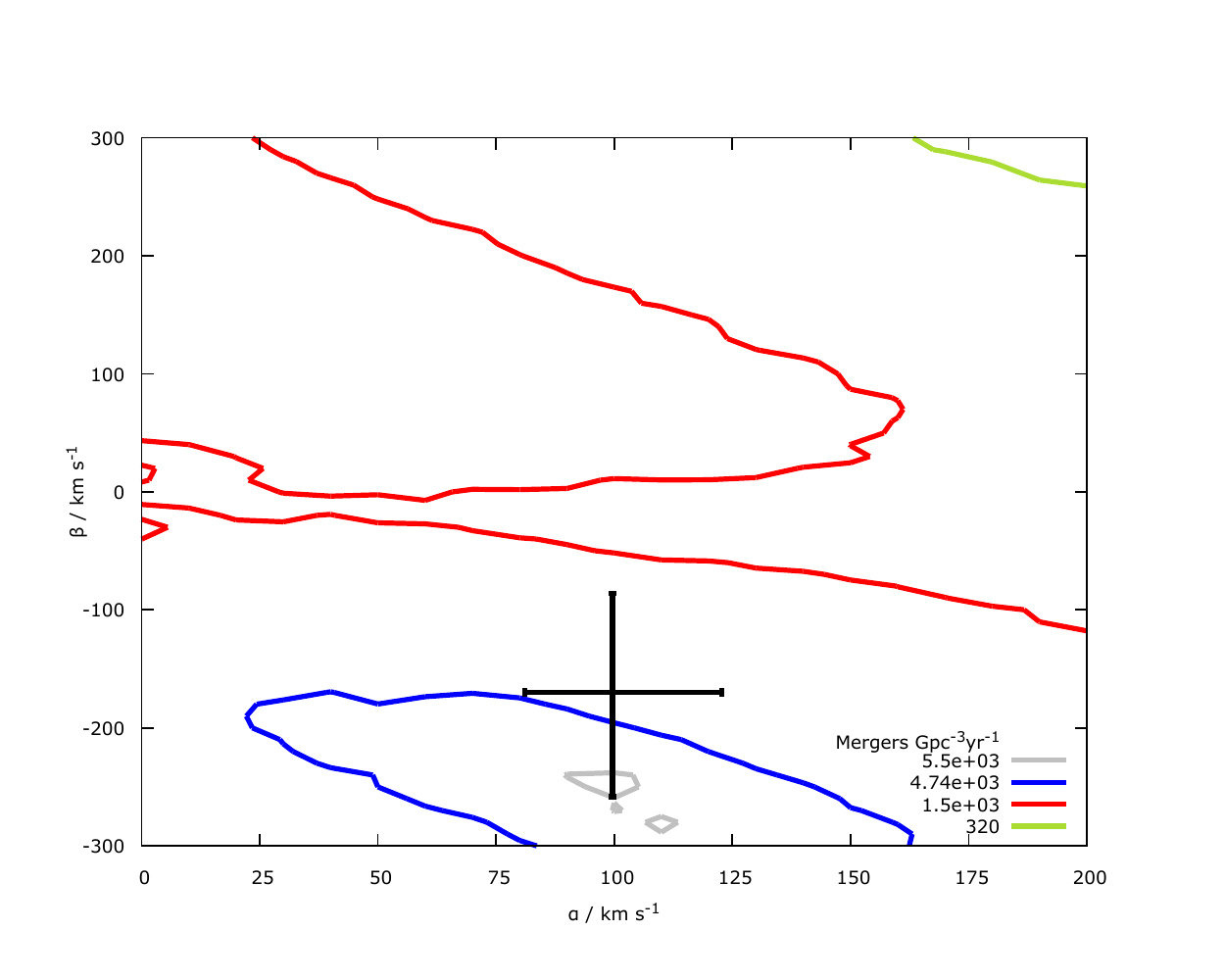}
	\caption{LH plot: Expected number of double-neutron-star (DNS) binary system mergers in a Milky Way equivalent galaxy Myr$^{-1}$ by $\alpha$ and $\beta$ combination. The black cross shows our best-fit of $\alpha=100$ and $\beta=-170$ (with error bars), giving 386 mergers.\\ RH plot: Expected number of double-neutron-star (DNS) binary system mergers Gpc$^{-3}$ yr$^{-1}$ by $\alpha$ and $\beta$ combination using a Milky way equivalent galaxy (MWEG) star formation rate (SFR) of 3.5~M$_\odot$ yr$^{-1}$ for 10~Gyrs. The green contour line shows the Laser Interferometer Gravitational-Wave Observatory (LIGO) lower limit of 320 Gpc$^{-3}$ yr$^{-1}$, the red line shows the calculated  LIGO detection rate of 1,540~Gpc$^{-3}$ yr$^{-1}$ and the blue line shows the LIGO upper limit of 4,740~Gpc$^{-3}$ yr$^{-1}$. The black cross shows our best-fit of $\alpha=100$ and $\beta=-170$ with error bars, giving 3,864 mergers.\\ }
	\label{fig:gpcmr}
\end{figure*}

{\bf{Merger Rate Calculations:}} Our first merger rate calculation is for the approximate DNS merger number for a Milky Way equivalent galaxy (MWEG). To enable a comparison of our rate to that of \cite{RN464}, for this calculation we use the Milky Way star formation rate of 3.5~M$_\odot$ yr$^{-1}$ for 10 Gyrs. We obtain this from our merger number for our instantaneous starburst of 10$^{6}$~M$_\odot$ by multiplying our rate by 3.5 $\times$ 10$^4$, (since our instantaneous burst of 10$^{6}$~M$_\odot$ is equivalent to 1~M$_\odot$ yr$^{-1}$ for 1~Myrs). We obtain our final time averaged rate by dividing by the number of years over which our mergers would occur (10 Gyrs) by 10$^{4}$ to convert to a Myr$^{-1}$ rate. Our predicted Milky Way merger numbers by $\alpha$ and $\beta$ combination are shown in the left-hand plot of Figure~\ref{fig:gpcmr} with the rate using our best-fit kick shown by the black cross. 

We find our best-fit $\alpha$ and $\beta$ combination provides a predicted rate an order of magnitude greater than the standard model rate of \cite{RN464}, predicting 386 mergers Myr$^{-1}$ compared to their 24~Myr$^{-1}$. However, when using a kick selected from the \cite{RN165} distribution we obtain an almost identical rate to \cite{RN464}, of 22 mergers Myr$^{-1}$ (see Table~\ref{table:3}). Such a close correlation is extraordinary given that different stellar models have been used. \cite{RN472} using the same \bpass models and a kick selected from the \cite{RN165} distribution obtain a merger rate of 52~Myr$^{-1}$ but we note that they assume NSs form up to 3~M$_\odot$ whereas our cutoff is 2~M$_\odot$ which most likely explains the higher merger rate.

\begin{figure*}
	\vspace{-7mm}
	\hspace{-88mm}
		\includegraphics[scale=0.65]{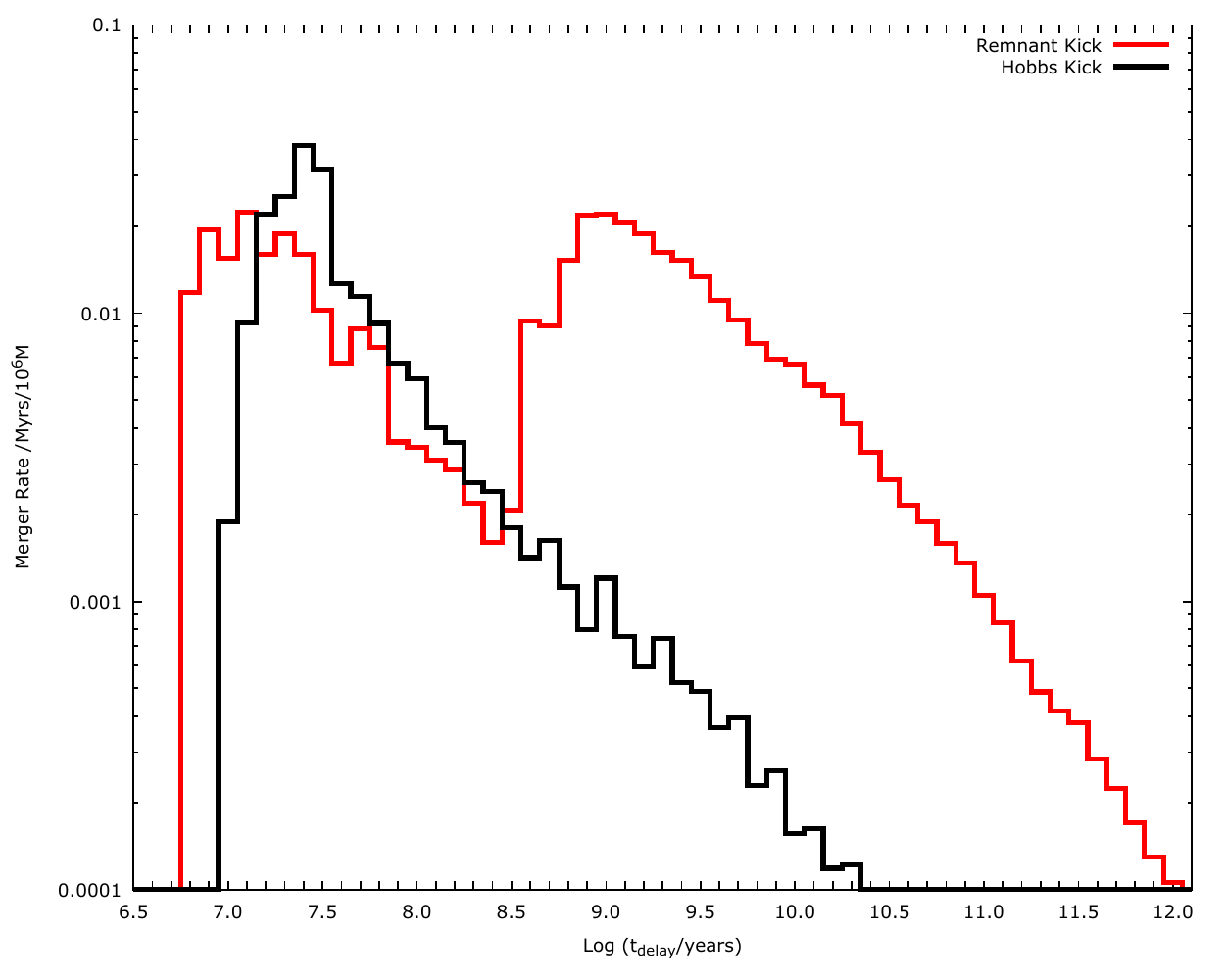}
\end{figure*}

For our second comparison we calculate the merger rate Gpc$^{-3}$ yr$^{-1}$ by $\alpha$ and $\beta$ combination. We start with our DNS merger number per 10$^{6}$~M$_\odot$ and convert this to the rate for a MWEG by multiplying by 3.5 $\times$ 10$^4$ and then dividing by 10$^{4}$ to convert to a Myr$^{-1}$ rate. We then multiply this rate by 10, as one merger Myrs$^{-1}$ in the Milky Way is equivilent  to 10 Gpc$^{-3}$ yr$^{-1}$. Our predicted Gpc$^{-3}$ yr$^{-1}$ rates are shown in the RH plot of Figure \ref{fig:gpcmr}. Our best-fit kick detection number of 3,864$^{+1,570}_{-2,371}$ Gpc$^{-3}$yr$^{-1}$ is shown by the black cross and while high, is within the upper limits of the LIGO DNS rate prediction.

 \begin{figure*}
	\vspace{0mm}
	\hspace{-88mm}
		\includegraphics[scale=0.75]{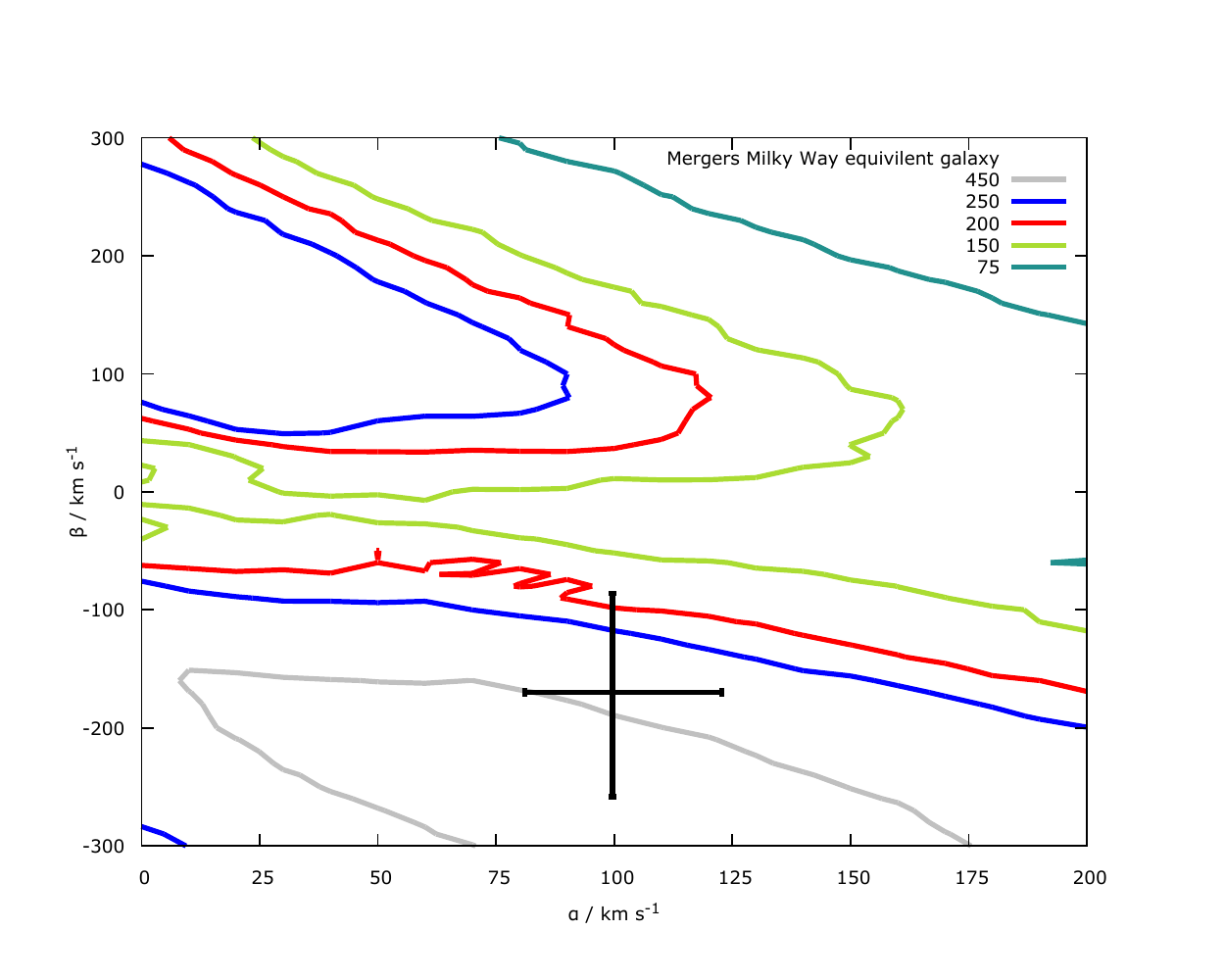}
\end{figure*}

\begin{table*}	
\caption{Comparison of best-fit kick and  \protect\cite{RN165} kicks for local merger rates: Local star formation rates (SFRs) assume 3.5 M$_\odot$ yr$^{-1}$ for 10 Gyrs: SFR by redshift after \protect\cite{RN487}}
\begin{center}
\begin{tabular}{ c c c c}
\hline\hline
 Merger rate calculation  & Best-fit kick & \protect\cite{RN165}& LIGO \\ 
 &($\alpha=100$ and $\beta=-170$)& kick  &predicted rate\\ 
 \hline\hline
Local DNS merger number Myr$^{-1}$ & 386 & 22 & -\\  
\hline
Local DNS merger number Gpc$^{-3}$ yr$^{-1}$ & 3864 & 220 & 320 - 4,740\\
\hline\hline
Local BH-BH merger number Gpc$^{-3}$ yr$^{-1}$ & 5 & 3.8 & 12 - 213\\
 \hline\hline
\end{tabular}
\\[1.5pt]
\textit{Note.} DNS=double neutron star; BH-BH=black-hole\textendash black-hole; Myr$^{-1}$=per million years; Gpc$^{-3}$ yr$^{-1}$=per cubic giga-parsec per year; LIGO= Laser Interferometer Gravitational-Wave Observatory 
\label{table:3}
\end{center}
\end{table*}

\section{Discussion}

With all comparisons of synthetic to observational datasets the challenge is to identify and mitigate selection effects in the observational data to ensure a relative comparison is being made while minimising manipulation and filtering of the data. For the runaways, we believe our inclusion of only those runaways with velocities which meet the widely accepted velocity cutoff of 28 km s$^{-1}$ meet this criteria. For DNS systems we believe this criteria was also meet by selecting those systems that would not merge in 50~Gyrs and whose periods do not exceed that of those observed. These selections were made to ensure that the number of binaries whose periods and eccentricities could have been modified by interactions were minimised and that we did not include systems whose periods meant their detection was unlikely.

Comparing the complete set of synthetic runaway velocities to the complete observational set of \cite{RN131} runaways using v$_{\rm pec}$, results in an overestimation of the runaway velocities. However, once the runaways with v$_{\rm pec}<$28 km s$^{-1}$ are removed our predicted velocities are in agreement with the observations. While we recognise that there will be a contribution from runaways dynamically ejected by the interaction of multiple star systems, once these have escaped from their OB clusters or associations, they are likely to have low velocities and hence be removed by our minimum velocity cutoff.

Similarly, while our synthetic DNS period distribution for systems that do not merge within a 50~Gyr period do not reproduce the observations, we record synthetic periods up to 10$^6$ days. Such long period systems are highly unlikely to be observed making their inclusion in comparisons unrealistic. Selecting synthetic DNS periods with the same maximum as those observed (log(P/days)$<$1.9), results in our synthetic periods providing a very good fit to the observations. 

\begin{table*}	
\caption{Best fit $\alpha$ and $\beta$ values and uncertainties using the isotropic kick for each observational dataset}
\begin{tabular}{c c c c c c c}
\hline\hline
Dataset& $\alpha$ &  $\beta$\\ 
&(${\rm km\,s^{-1}}$)&(${\rm km\,s^{-1}}$)\\
\hline\hline
Single NS velocity (2D) & $100^{+30}_{-20}$ & $-170^{+100}_{-100}$\\
\hline
Runaways (2D velocity - All)&0&0\\
Runaways (2D velocity - $v_{\rm pec}$ $<$ 28 ${\rm km\,s^{-1}}$)&110&-20\\
\hline
DNS period  (All periods)&0&-630\\
DNS period (Period$<$1.9 log days)&20&20\\
\hline
DNS eccentricity (All periods)&190&-60\\
DNS eccentricity (Period$<$1.9 log days)&190&-60\\
\hline\hline
\label{table:2}
\end{tabular}
\\[1.5pt]
\textit{Note.} NS=neutron star; DNS=double neutron star; 2D=two-dimensional; $v_{pec}$=peculiar or 3D velocity
\end{table*}

Our only problematic comparison is that of the DNS eccentricities where there is some tension between our best-fit $\alpha$ and $\beta$ values and the DNS eccentricity observations. While the uncertainties for our best-fit $\alpha$ and $\beta$ values do cross into the region of the observational critical KS D$_{\alpha}$ value, generally our best-fit values result in systems with much higher eccentricities than those observed. We suspect this poor fit is most likely a result of our circularisation of the NS-companion star orbit prior to the secondary supernovae. If there were pre-existing eccentricities the second kick would most likely reduce some eccentricities and increase others, resulting in some of the higher eccentricity systems being disrupted. The net effect is expected to be a reduction in the overall eccentricity distribution as observed.

Interestingly, we find using our best-fit $\alpha$ and $\beta$ values that some NSs experience a negative kick as a result of the supernova, i.e. the neutron star gains a velocity {\it toward} the largest ejecta mass rather than away from it. While this would seem to contradict the findings of \cite{RN474} these NSs would almost certainly be from low-mass or stripped binary companions and not be created by single star evolutionary pathways. As a result we would not expect to observe single NSs co-moving with the most massive ejecta mass, which would appear to be the case. 

Using our best-fit $\alpha$ and $\beta$ values, our estimated DNS merger rate of 3,864~Gpc$^{-3}$yr$^{-1}$ is just within the current upper limit for the LIGO DNS number estimate of 1,540$^{+3,200}_{-1,220}$~Gpc$^{-3}$ yr$^{-1}$ \citep{RN488}. Using the same best-fit $\alpha$ and $\beta$ values we replicate the work of \cite{RN464} and calculate the corresponding BH-BH merger number, and find a merger number of 5$^{+40}_{-1}$ Gpc$^{-3}$yr$^{-1}$, compared to the lower-limit of the current LIGO estimate of 12-213 Gpc$^{-3}$ yr$^{-1}$ \citep{RN489}. The primary reason our prediction is not close to the LIGO measured merger rate is that we limit ourselves to solar metallicity where all other simulations show that the rate of these mergers is dominated by compact remnants from lower metallicities \citep{Belczynski:2016aa,RN472,RN459}.

We stress that these are approximate rates which only consider the local SFR and do not consider the time evolution of metallicity. While these factors may not be significant for NS formation rates, the impact of metallicity becomes more important when calculating BH formation rates and hence our BH-BH merger rates need to be interpreted with caution. 

We note that using the same \bpass models and a kick selected from the \cite{RN165} distribution, \cite{RN472} obtain a BH-BH merger rate at solar metallicity of 0.82~Myrs$^{-1}$, while ours is slightly lower at 0.38~Myrs$^{-1}$. The difference is most likely a result of their kick angles not being modified by the $\arccos$ function and so their kicks are more pole-centred and as a result fewer systems are disrupted. 

Using our original best-fit kick values of $\alpha=70$ and $\beta=120$ \cite{RN464} find merger rates for DNS binaries of 85.5~Myr$^{-1}$ and 244.8~Gpc$^{-3}$yr$^{-1}$ which is below the current LIGO estimate. Replicating this calculation in our code we find very similar rates of 78~Myr$^{-1}$ and 273~Gpc$^{-3}$yr$^{-1}$. 

Probably the most significant consequence of the negative $\beta$ value is not the presence of negative kicks but the overall reduction in kick magnitudes especially for NSs formed in low-mass or stripped supernovae. These low-mass events are more common in the secondary supernovae with the obvious consequence that binary systems where the secondaries experience mass-stripping have smaller kicks according to our proposed kick model, and are much more likely to survive the supernovae. This seems to support the research of \cite{RN386} who suggest there are at least two kick distributions. The result is an increased occurrence of DNSs and hence in DNS merger rates. However, because of their necessary larger final masses, secondary progenitors forming BHs will have much larger conservation of momentum kicks and hence be less affected by this negative $\beta$ value. We suspect this explains why the increase in the DNS merger rate is not accompanied by a corresponding increase in the BH-BH merger rate and why our BH-BH merger rate does not exceed that predicted by the LIGO collaboration.

We also note that if the $\beta$ value is a gravitational effect, it is not a constant but related to the level of mass asymmetry in the ejecta. We have not investigated a mass dependence in our $\beta$ value and this presents a possible avenue for future research.



For both our synthetic DNS eccentricity and period distributions we did expect somewhat higher eccentricities and longer periods than the observational dataset. This is because our synthetic distributions are calculated at the time of the formation of the DNS binary systems, whereas even selecting observational systems that do not merge within a 50~Gyr period, will result in the inclusion of systems that may have had a significant time to partially circularise, resulting in reduced eccentricities, and to a lesser degree periods. 




\section{Conclusion}
In light of comments by \cite{RN432}, we have modified and extended our \textsc{reaper} code and recalculated our best-fit $\alpha$ and $\beta$ values. These are now $\alpha=100$ and $\beta=-170$. Through basic kinematics, we have shown these $\alpha$ and $\beta$ values are plausible using standard physical processes namely conservation of momentum and gravitational interaction. While we in no way claim this as the definitive solution to providing the magnitude of the supernova kick, we believe these values provide a valuable link between the physical properties of the progenitors and the resulting supernova kicks. 

We also accept that our observational comparisons are by no means conclusive proof of the validity of the conservation of momentum kick, 
but highlight that recent observations by \cite{RN242} and \cite{RN474} do appear to support this type of mechanism as the main contributor to the NS kick. In particular, the recent research by \cite{RN474} on six young core-collapse supernova remnants showing NS motions roughly opposite and proportional to that of the intermediate-mass element ejecta, seem to support the conservation of momentum kick as the main source of the remnant velocities.

While there is some tension between the comparison of our best-fit kick values and the observations for DNS eccentricities, in all comparisons our best-fit kick values are \textit{not} ruled out by any of the two-sample KS tests. In fact if our ``toy'' model is correct, our $\beta$ value is most likely dependant on the ejecta mass, and the omission of this relationship may explain why the eccentricities are in tension with the observations. In the case of the merger number calculations, our best-fit kick values are in agreement with the predicted LIGO DNS merger rates, and the same $\alpha$ and $\beta$ values predict BH-BH merger rates very close to those predicted by the LIGO collaboration. Our merger rate calculations are also broadly in line with those predicted by other researchers when using a kick randomly selected from the \cite{RN165} distribution.

Finally, we reiterate that the most fundamental advantage of our kick prescription is that it provides a link between the remnant velocity and the physical properties of the progenitor star. If we are correct and such a kick relationship exists with best-fit values of the order we have identified, we would expect significantly more DNS mergers to be detected in the next LIGO detection run. 

\section*{Acknowledgements}

JCB and JJE acknowledge the support from the University of Auckland.

We would like to thank Thomas Janka, Noam Soker and the anonymous referee for their valuable comments and suggestions to improve this paper.

The authors wish to acknowledge the contribution of the NeSI high-performance computing facilities and the staff at the Centre for eResearch at the University of Auckland. New Zealand's national facilities are provided by the New Zealand eScience Infrastructure (NeSI) and funded jointly by NeSI's collaborator institutions and through the Ministry of Business, Innovation and Employment's Infrastructure programme. URL: http://www.nesi.org.nz
\addtocontents{toc}{\protect\vspace*{\baselineskip}}
\bibliography{RefMASTER.bib}
\begin{appendices}
\onecolumn
\begin{figure}
\section{Supernovae Outcomes from reaper}\label{A1}
\centering
\vspace{0mm}
		\includegraphics[scale=0.75]{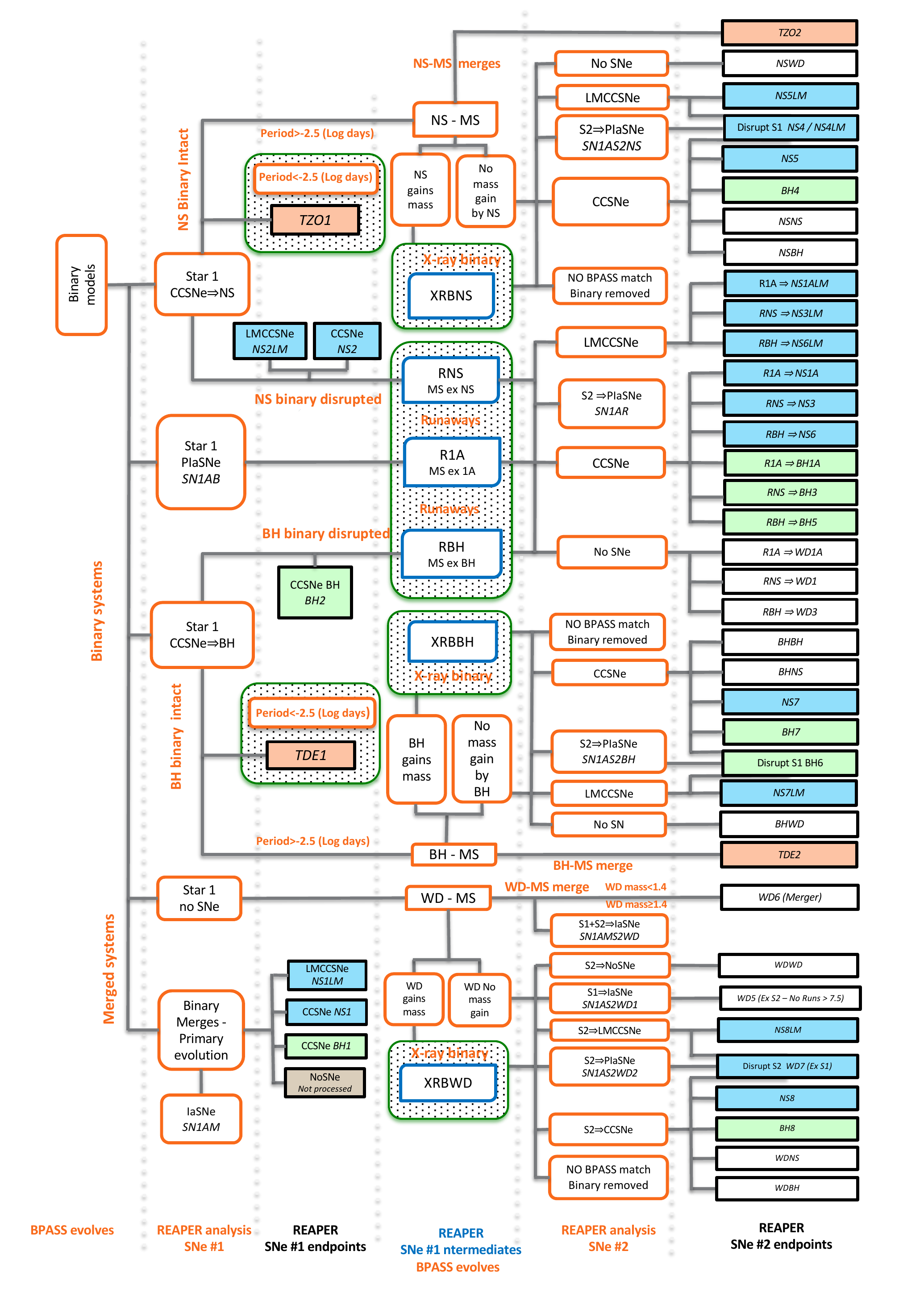}
	\vspace{-0mm}
	\caption{Output map for \textsc{reaper} analysis of \textsc{bpass}v2 evolutionary models}
	\label{fig:A1}
	Key: SNe=supernova: CC=core collapse: LM=low-mass: PIa=prompt Type Ia supernova: MS=main sequence star: R=runaway star: WD=white dwarf: NS=neutron star: BH=black hole: TZO=Thorne Zytkow object: TDE=tidal disruption event: XRB=X-ray binary
\end{figure}

\onecolumn
\begin{figure}
\section{Single Pulsar 2D Velocity Distributions and Best-fit $\alpha$ and $\beta$ combinations}\label{A2}
The six cumulative 2D pulsar velocity distributions tested and their best-fit $\alpha$ and $\beta$ combination for each are listed below with the corresponding cumulative 2D velocity distributions shown in Figure \ref{fig:A2}.
\begin{itemize}
\item Subset of \cite{RN165} pulsars, characteristic ages \textless3 Myrs. $\alpha=100$ and $\beta=-170$
\item Subset of \cite{RN165} pulsars, characteristic ages \textless7 Myrs. $\alpha=120$ and $\beta=-280$
\item Subset of \cite{RN165} pulsars, characteristic ages \textless3 Myrs, EDM distance measurements. $\alpha=110$ and $\beta=-240$
\item Subset of \cite{RN165} pulsars, characteristic ages \textless7 Myrs, EDM distance measurements. $\alpha=110$ and $\beta=-260$
\item ATNF pulsars, characteristic ages \textless3 Myrs. $\alpha=120$ and $\beta=-230$
\item ATNF pulsars, characteristic ages \textless7 Myrs. $\alpha=120$ and $\beta=-260$
\end{itemize}
\vspace{10mm}
	\centering
		\includegraphics[scale=1.0]{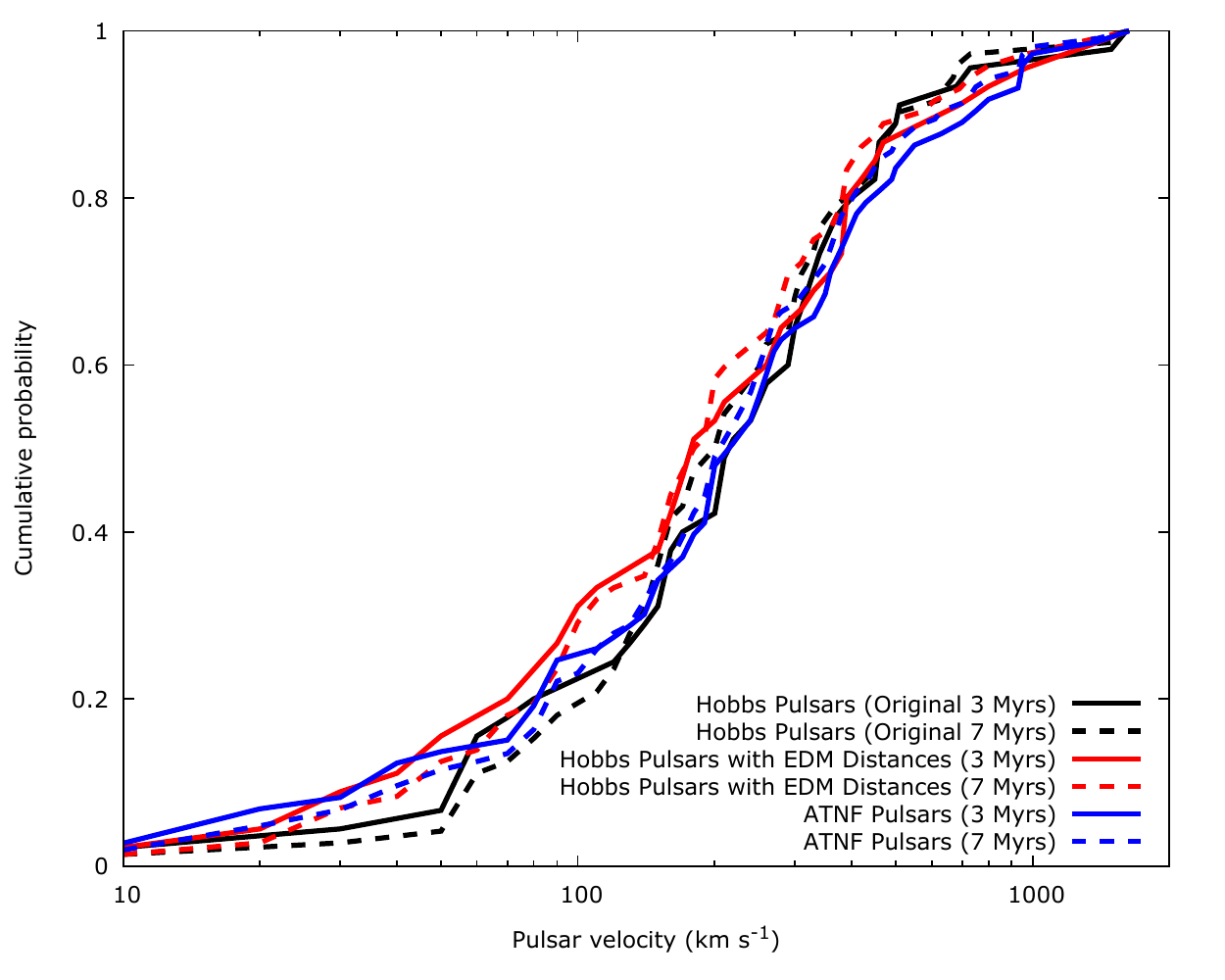}
	\caption{The cumulative 2D pulsar velocity distributions for the six datasets tested.}
	\label{fig:A2}
	Key: EDM=electron density measurement: ATNF=Australia telescope national facility 
\end{figure}	
As expected, uncertainties in distance and proper motion measurements have a much larger effect on the low velocity pulsars with velocities above 200 km s$^{-1}$ virtually unchanged.
\end{appendices}
\end{document}